\documentclass[review, english, 12pt]{elsarticle}
\usepackage[margin=2.5cm]{geometry}
\biboptions{sort&compress}
\RequirePackage{amsthm, amssymb, amsmath, amsfonts}

\usepackage[x11names, usenames, dvipsnames, svgnames, table]{xcolor}
\usepackage{textgreek} 
\usepackage{verbatim}
\usepackage{enumitem}
\RequirePackage[skip=2pt, labelfont=bf]{caption}
\RequirePackage[skip=1.75pt, font=footnotesize, labelfont=sf, textfont=sf]{subcaption}
\usepackage{textcomp}
\usepackage{tabularx}
\usepackage{graphicx}
\usepackage{xtab, booktabs} 
\usepackage{lscape}
\usepackage{longtable}
\usepackage{adjustbox}
\usepackage{setspace}
\usepackage{multirow}
\usepackage{multicol}
\usepackage{tcolorbox}
\usepackage{hhline}
\usepackage{colortbl}
\usepackage{rotating}
\usepackage[normalem]{ulem}
\usepackage{bm}
\usepackage{nicefrac}  
\usepackage{microtype}  
\usepackage{easylist}
\usepackage[hyphens]{url}

\usepackage[colorlinks]{hyperref}
 \AtBeginDocument{%
     \hypersetup{colorlinks=true, citecolor=Purple, linkcolor=Blue, urlcolor=NavyBlue, 
} }

 \RequirePackage{algorithm}
 \RequirePackage{algorithmic}
 \usepackage[algo2e, ruled, linesnumbered, resetcount]{algorithm2e}
 \def\HiLi{\leavevmode\rlap{\hbox to \hsize{\color{yellow!50}\leaders\hrule height .80\baselineskip depth .8ex\hfill}}}

 \journal{Applied Soft Computing}

\begin{document}
\newcommand*{\everymodeprime}{\ensuremath{\prime}}
 \begin{frontmatter}
\title{An ensemble meta-estimator to predict source code testability}

\author[mainaddress]{Morteza Zakeri-Nasrabadi}
\ead{morteza\_zakeri@comp.iust.ac.ir}

\author[mainaddress]{Saeed Parsa\corref{correspondingauthor}}
\cortext[correspondingauthor]{Corresponding author}
\ead{parsa@iust.ac.ir}
\address[mainaddress]{School of Computer Engineering, Iran University of Science and Technology, Tehran, Iran.}

\begin{abstract}
Unlike most other software quality attributes, testability cannot be evaluated solely based on the characteristics of the source code. The effectiveness of the test suite and the budget assigned to the test highly impact the testability of the code under test. The size of a test suite determines the test effort and cost, while the coverage measure indicates the test effectiveness. Therefore, testability can be measured based on the coverage and number of test cases provided by a test suite, considering the test budget.
This paper offers a new equation to estimate testability regarding the size and coverage of a given test suite. The equation has been used to label 23,000 classes belonging to 110 Java projects with their testability measure. The labeled classes were vectorized using 262 metrics. The labeled vectors were fed into a family of supervised machine learning algorithms, regression, to predict testability in terms of the source code metrics.
Regression models predicted testability with an R\textsuperscript{2} of 0.68 and a mean squared error of 0.03, suitable in practice. Fifteen software metrics highly affecting testability prediction were identified using a feature importance analysis technique on the learned model. The proposed models have improved mean absolute error by 38\% due to utilizing new criteria, metrics, and data compared with the relevant study on predicting branch coverage as a test criterion. As an application of testability prediction, it is demonstrated that automated refactoring of 42 smelly Java classes targeted at improving the 15 influential software metrics could elevate their testability by an average of 86.87\%.
\end{abstract}
\end{frontmatter}

\section*{Code metadata}
\noindent
Permanent link to reproducible Capsule: \href{https://doi.org/10.24433/CO.9048896.v1}{https://doi.org/10.24433/CO.9048896.v1}.

\section{Introduction}\label{sec:introduction}
Software testing is an undecidable problem \cite{Dijkstra1972, Ammann2016}, while testability is a decidable property of software. If testing were decidable, there would be no reason for testability. So far, software testability has been primarily measured in terms of source code metrics \cite{Khan2009, MuhammadRabeeShaheen2014, Suri2015}. 
Cyclomatic complexity (CC) for a method \cite{Cohen1989}, weighted methods per class (WMC) \cite{Chidamber1994}, lack of cohesion of a method (LCOM) \cite{Cohen1989}, tight class cohesion (TCC) \cite{Bieman1995}, and loose class cohesion (LCC) \cite{Bieman1995} are examples of source code metrics mostly applied to measure testability. Controllability and observability are the other known metrics used to measure testability \cite{Garousi2019, Sharma2018}. All of These metrics are computed statically without considering test effectiveness and effort. However, according to the standards \cite{IEEEStd610.12-1990, ISOandIEC2011}, the two significant factors affecting the testability are the test effectiveness and efficiency, which cannot be computed without considering the runtime behavior.

ISO/IEC 25010:2011 standard \cite{ISOandIEC2011} defines testability as the" degree of effectiveness and efficiency with which test criteria can be established for a system, product or component and tests can be performed to determine whether those criteria have been met." Based on this definition, test efficiency and effectiveness are assumed to be contingent upon testability and vice versa. Recently, there have been some considerations for test effectiveness and efficiency as the two main ingredients to measure testability. 
For instance, a recent attempt by Terragni et al. assumes a direct relation between testability and test effort when formulating testability \cite{Terragni2020}. 
 
 Several complexity metrics have been used to estimate testability solely from a test effort point of view  \cite{Cohen1989, Bieman1995, Bruntink2004, Bruntink2006, BadriLinda2011, BadriMourad2012}. These metrics can take any value that does not precisely indicate the testability measure. Moreover, recent studies show a moderate correlation between software metrics and testability \cite{BadriLinda2011, BadriMourad2012}. The moderate correlation implies that metrics do not entirely measure testability \cite{Oluwatosin2020}, and there are factors involved in developing test cases that software metrics cannot capture. 
 Indeed, the correlation is measured between the metrics and testability regarding the test effort. This paper shows that the correlation could be strong enough, provided that besides the test effort, the test effectiveness was considered. 
 
 This paper offers a \emph{mathematical model} to precisely compute testability in terms of the tests' effectiveness and efficiency. The effectiveness is subject to the test coverage, while the efficiency depends on the inverse of the effort made to establish the test criteria: the higher testability, the less effort required to test effectively. The mathematical model computes testability using the runtime information collected by an automated test data generator, EvoSuite \cite{Panichella2020}, for Java classes. 
 The test effectiveness and effort used by the mathematical model may be affected by tester skills and budgets, which are unknown for manual tests. However, testability is an inherent attribute of software. 
Therefore, it is preferred to generate test data automatically. 
 In this way, fair testing conditions are provided for all the classes under test to avoid any external factor that may affect testability. Otherwise, human factors such as the tester skills should be added to the suggested testability model.
 
 The two significant downsides of the mathematical model are the possibility of lengthy execution times and the reliance on the executable code. Therefore, once and for all, the mathematical model was used to calculate the testability of 23,000 Java classes.
The classes were then vectorized using 264 software metrics. The vectors were then labeled with their corresponding testability computed by the mathematical model. The labeled vectors were used as samples to train an ensemble of meta-estimators built upon three base regressors to predict class testability without running the program under test. The learned model predicts testability statically in terms of source code metrics.  

The authors have already introduced a model to predict test effectiveness in terms of a new metric called \emph{Coverageability} \cite{Zakeri2021}. Coverageability indicates the extent to which a given source code may be covered with test data generated automatically. Coverageability, indeed, can be considered as a measure of test effectiveness. Test efficiency is another factor affecting testability. Test efficiency opposes its effectiveness. Therefore, a compromise should be struck between the effectiveness and efficiency of the test to ensure reasonable testability. The proposed approach in this paper considers both the test effectiveness and efficiency factors to compute testability. 

Practically, as the number of statistical features increases, the system's accuracy improves \cite{Zakeri2021}. The accuracy of the proposed model was improved when using statistical (minimum, maximum, sum, mean, and standard deviation) values of the metrics. This way, the number of metrics used to train the model increased to 262. A feature importance analysis technique \cite{Breiman2001} was applied, computing the impact of each metric on testability prediction to support the interpretability \cite{Carvalho2019} of the learned model. All the source code metrics were then ranked and sorted according to the model sensitivity to each metric. Finally, a set of refactoring operations that highly change influential metrics was selected and applied to improve the class testability. 

 The effect of refactoring on source code metrics and, subsequently, the impact of the source code metrics on the testability of the unit under test has enabled estimation of the impact of refactorings on the testability. The proposed model has made it possible to develop testable code by measuring testability before and after refactoring. Frequent refactoring followed by testability measurement leads to efficient and effective tests, reducing testing costs. Moreover, as testability improves, some quality attributes, including reusability, functionality, extensibility, and modularity.
In summary, the significant contributions of this research to the software testability literature are as follows:

\begin{enumerate}
        \item{To establish a mathematical model that relates test efficiency and effectiveness with software testability.}
        \item{To predict testability value using source code metrics without any need to run the program.}
        \item{To designate the most influential source code metrics affecting testability prediction.}
        \item{To provide the opportunity to measure and improve testability while developing code.}
\end{enumerate}

The experimental results on 110 software projects show that the proposed model could learn and predict the testability of 23,000 Java classes with an R\textsuperscript{2} of 0.68 and a mean squared error of 0.03. Moreover, the experiments with automatically refactoring 42 Java classes demonstrate relatively significant improvement in source code metrics, as a result of which, on average, testability was enhanced by 86.87\%. In addition, other quality attributes, including reusability, functionality, extendability, and modularity, were improved.

The remaining parts of this paper are organized as follows: Section \ref{sec:background-and-related-work} discusses the background and related works. Section \ref{sec:methodology} describes a new testability definition and proposes a methodology for measuring the source code testability.
Experimental evaluation results are given in Section \ref{sec:experiments}. 
Section \ref{sec:threats-to-validity} discusses threats to validity.
The conclusion and feature work are discussed in Section \ref{sec:conclusion}.

\section{Related work}\label{sec:background-and-related-work}
\noindent
There are several definitions for software testability, none of which specify how to measure testability, which has led researchers to develop different approaches \cite{Garousi2019}.
The recent survey by Garousi et al. \cite{Garousi2019} has listed more than 30 definitions for software testability. The IEEE, ISO, and IEC standards propose seven of these definitions. According to their survey, the most common definitions are as follows:

\begin{enumerate}
    \item{
IEEE standard 610.12-1990 \cite{IEEE1990}: "the degree to which a system or component facilitates the establishment of test criteria and the performance of tests to determine whether those criteria have been met; the degree to which a requirement is stated in terms that permit the establishment of test criteria and performance of tests to determine whether those criteria have been met."
}

\item{
 ISO standard 9126-1: 2001 \cite{ISO2001}: "attributes of software that bear on the effort needed to validate the software product."
}

\item{
	ISO/IEC standard 25010:2011 \cite{ISOandIEC2011}: "degree of effectiveness and efficiency with which test criteria can be established for a system, product or component and tests can be performed to determine whether those criteria have been met."
}
\end{enumerate}

The most recent definition proposed by ISO/IEC standard 25010:2011 \cite{ISOandIEC2011} is considered in this paper to establish a novel testability prediction model.
Many approaches provide mathematical models for calculating testability \cite{Bruntink2006, Khan2009, Toure2018, BadriMourad2019}. Their results can not be generalized because they investigate at most eight software projects \cite{Terragni2020}. They also use manually generated test cases which could be a reason for the low number of projects. 

Bruntink and van Deursen \cite{Bruntink2004, Bruntink2006} have investigated the correlation between C\&K metrics and test effort in terms of line of code per class (dLOCC) and the number of test cases (dNOTC). However, they have not considered the test budget and test effectiveness factors in computing the correlation. Moreover, they have only used five Java projects to evaluate their results.
 Badri et al. \cite{BadriLinda2011} concluded a moderate correlation between cohesion metrics and test effort using only two software systems. The moderate correlation implies that these metrics do not entirely measure the test effort.  
Toure et al. \cite{Toure2018} introduced a more complex metric called Quality Assurance Indicator (Qi) based on the concept of control call graph (CCG). However, they have not considered any runtime information, including test adequacy criteria, in their measurement.
Badri et al. \cite{BadriMourad2019} used linear regression analysis, and five machine learning algorithms have been used to develop explanatory models. However, they did not consider the metric impacts on test adequacy criteria.

Controllability and observability concepts have been explored broadly for measuring software testability \cite{Garousi2019, Sharma2018}. 
There are different definitions for observability and controllability properties in software systems. In general, observability determines how easy it is to observe the behavior of a program (part of the program) in terms of its outputs, effects on the environment, and other hardware and software components \cite{Voas1995}. It emphasizes the ease of observing program outputs. For instance, passing argument by reference reduces the program observability since the inputs may change by the body of the called method. 
Controllability is the degree to which the state of the component under test as required for testing can be controlled. It focuses on the ease of producing a specified output from the specified input \cite{Binder1994}. For instance, the polymorphic method reduces the controllability since different paths in the program are executed based on the type of input that inference implicitly.  

Observability and controllability concepts are mainly used to assess the testability of hardware components. The adaption of these concepts to software systems has been performed by many authors \cite{Garousi2019}. However, there are very limited practical implementations and empirical studies on measuring software testability based on the observability and controllability metrics. 
COTT \cite{Goel2008} is a framework to help the software tester observe the internal behavior and control the difficult-to-achieve states of the software under test (SUT). It needs the SUT to be instrumented and executed for collecting the controllability and observability information, exhausting for large source codes. 

Runtime testability is defined as the maximum code coverage achieved when running tests  \cite{Gonzalez-Sanchez2010}.
Code coverage \cite{Ammann2016} achieved by a test suite provides a sound indication of test effectiveness and quality \cite{Terragni2020}.
Salahirad et al. \cite{Salahirad2019} have shown that branch coverage is the most influential criterion for using in the fitness functions of search-based test data generation tools. They have concluded that it is more difficult to automatically generate tests for less visible class methods, \textit{e.g.}, the private method.
Testability, therefore, has severe impacts on the result of automated testing. The observation was that the relationship between testability and code coverage had not been studied when applying automated testing.
 Ma et al. \cite{Ma2017} studied the impact of code visibility metrics on code coverage in manual and automated testing.
Code visibility refers to the accessibility of entities from other entities in the program. In object-oriented programming, access modifiers such as private, protected, and public provides code visibility and information hiding facilities.
 Results demonstrate that \emph{developer-written} tests are insensitive to code visibility; however, \emph{automatically generated} tests yield relatively lower code coverage on less visible code. Automated testing obstacles and issues could be explored by establishing a relationship between source code metrics and testability.
 
The early works to estimate code coverage are proposed by Daniel and Boshernitsan \cite{Daniel2008}. They created a decision tree classifier for Java projects to predict the coverage level of the method under test before testing with an automatic test data generation tool. However, their trained classifier could estimate the coverage level in two classes of high and low only. Therefore, their proposed model is not suitable for discriminating programs based on their testability values.
 
 Ferrer et al. \cite{Ferrer2013} have introduced a new complexity measure, branch coverage expectation (BCE), to estimate the number of test cases required to achieve full branch coverage. To this aim, they have transformed the program control graph (CFG) into the first-order Markov chain \cite{Kobayashi2011}. They hard-coded the probability of transitions between basic blocks represented as states in the Markov Chain and used the average stationary probability of all basic blocks to compute BCE. They concluded that traditional metrics do not estimate the coverage obtained with test-data generation tools compared to BCE for the corresponding code snippet. 
 However, computing the stationary distribution for all basic blocks in large programs is time-consuming and may result in space state explosion. In addition, having no sound justification for the probabilities assigned to the transitions in the Markov Chain threatens the accuracy of the results.
 
The code coverage level is practically measured in terms of statement or branch coverage as continuous variables. Regression learning is preferred most when the target variables are continuous. Grano et al. \cite{Grano2019} have created and compared four regression models, including Huber regression \cite{Hampel2011}, support vector regression \cite{Chih-Chung2011}, multilayer perceptron \cite{Goodfellow2016}, and random forest regression \cite{Breiman2001}, to predict the branch coverage level of the class under test based on 79 source code metrics. 
Despite using various source code metrics, data samples, and configuration parameters to tune the models, they have reported a relatively low prediction performance. 
The empirical analysis of the interrelationship between branch coverage and source code metrics  indicates that branch coverage is poorly correlated with source code metrics, which leads to low performance when machine learning is used for prediction. This observation is supported by the fact that test efforts directly impact the branch coverage in addition to the source code attributes, measured in terms of the metrics.

In recent work, Terragni et al. \cite{Terragni2020} normalized test effort metrics (number of test cases, number of assertions, and lines of test codes) with test adequacy metrics (line coverage, branch coverage, and mutation score) and showed that this normalization boosts the correlation between test effort and test adequacy metrics. Their approach can only estimate test effort based on only one metric at a time. More importantly, like other related works, they use existing human-written tests as benchmarks. Indeed, as an inherent feature, human factors should not affect testability.

So far, machine learning approaches have been applied to different aspects of software testing and debugging \cite{Noorian2011}, including test data generation \cite{Zakeri2020}, fault prediction \cite{Abdi2015, Shi2020, Mesquita2016}, and fault localization \cite{Maru2019, Dutta2021}.
Mesquita et al. \cite{Mesquita2016} have used the extreme learning machine (ELM) algorithm to classify source code modules as faulty and nonfaulty with a reject option using 17 source code metrics. If the faulty and nonfaulty classes are almost equally probable, the classifier rejects the sample instead of choosing a label. 
The source code metrics introduced in this article are 262, which provide a relatively good feature space for any machine learning technique applied to source code analysis. For sure, by increasing the 17 metrics to 262, the accuracy of the fault prediction models will also improve.

Shi et al. \cite{Shi2020} have proposed a new source code representation method, PathPair2Vec, based on path pairs in abstract syntax trees. Their approach converts the source code of a method to a fixed-length-vector which is then used as a feature vector for the code defect prediction task. PathPair2Vec \cite{Shi2020} outperforms code2vec \cite{Alon20191} and code2seq \cite{Alon2018} approaches which are similar code embedding methods. 
However, the automatic code embedding approaches are computationally intensive and require many code samples to train properly. 
In contrast, this paper proposes a lightweight approach to improve feature space by systematically adding new source code metrics. 

Xiao et al. \cite{Xiao2020} have incorporated test efforts in software fault detection and correction process. The authors have modeled the software testing process by a tri-process combining testing effort, fault detection process (FDP), and fault correction process (FCP) in Consecutive testing periods. Different neural network architectures have been used to predict the next triplet of the testing effort, FDP, and FCP, based on a sequence of previous ones. The testing effort in their experiments is expressed as the CPU hours or the total number of weeks consumed in the manual testing and correction process. However, this type of effort is highly influenced by tester skills. This paper uses automatic test data generation to keep the tester's skill and other factors affecting test results, such as the time budget fixed for all projects.

Dutta et al. \cite{Dutta2021} have presented a hybrid approach, Combi-FL, for effective fault localization by combining neural network, spectrum-based, and mutation-based fault localization techniques. Four out of eight techniques used in Combi-FL are based on neural networks, which emphasize the effectiveness of learning-based methods. 

The methodology proposed in the next section leverages supervised learning techniques to build a testability prediction model based on the standard definition of testability \cite{ISOandIEC2011}. It applies a well-known test data generation tool, EvoSuite \cite{Arcuri2016}, to 110 Java open-source projects \cite{Fraser2014} to generate test suites and collect the code coverage levels by the generated tests. EvoSuite is selected because it is a state-of-the-art tool \cite{Panichella2020}, attempting to maximize many test adequacy criteria by employing evolutionary algorithms compared to the pure random test data generators \cite{PachecoE2007Poster}.

\section{Methodology}\label{sec:methodology}
\noindent
The proposed testability measurement methodology follows the system and software quality models standard, ISO/IEC25010: 2011 \cite{ISOandIEC2011}, emphasizing two aspects of \emph{test efficiency} and \emph{test effectiveness} required to establish and satisfy given test criteria. 
Section \ref{sec:formalism} offers a mathematical model considering these two aspects to compute testability. 
The mathematical model is used to label samples used for building the machine learning model, described in Section \ref{sec:testability-measurement-framework}.
 
\subsection{Testability formal definition}\label{sec:formalism}
\noindent 
For a given class, $X$, testability, $T(X)$, is defined as the product of its test effectiveness, $T_Q\left(X\right)$, and test efficiency, $T_P\left(X\right)$:

\begin{equation}\label{eq:1}
T\left(X\right) = T_Q\left(X\right) \times T_P\left(X\right)
\end{equation}

The test effectiveness, $T_Q\left(X\right)$, of the class, $X$, is computed as the average of different coverage criteria, $Criteria$, considered for a given test suite:
\begin{equation}\label{eq:2}
T_Q\left(X\right)\ =\ \frac{1}{|Criteria|}\sum_{c \in Criteria}{c^{level}(X)}
\end{equation}
where $c^{level}(X)$ denotes the covered level of the given criterion, $c$. The test efficiency, $T_P\left(X\right)$, is considered as the reciprocal of test effort, $T_E\left(X\right)$:
\begin{equation}\label{eq:3}
T_P\left(X\right)=\frac{1}{T_E\left(X\right)}
\end{equation}

The main factor used to measure test effort in software testing literature is the test suite size \cite{Terragni2020}. The empirical observations in this article show that as the number of tests increases, the growth rate of the code coverage decreases. 
Suppose there are ten paths in a given source code. The objective of a test data generator is to generate influential test data not overlapping with the existing one. Therefore, when generating the first test data, the probability of generating test data that covers a path that is already covered by the existing test data is zero. When generating the second test data, the probability will be 0.1, and when generating the ninth test data, the probability will be 0.9. Therefore, as the number of influential test data affecting the coverage increases, the probability of generating redundant test data not affecting the coverage increases.

On the other hand, generating more \emph{influential tests} with the same time budget results in higher testability. 
An influential test is a test that strictly increases the code coverage. Therefore, for the class, $X$, test effort, $T_E\left(X\right)$, can be measured as follows:

\begin{equation}\label{eq:4}
T_E\left(X\right) = \left(1+\omega\right)^{\left\lceil\frac{|\tau\left(X,c\right)|}{NOM(X)}\right\rceil-1}
\end{equation}
where $\tau\left(X,c\right)$ is the minimized test suite containing influential tests normalized by the number of methods, $NOM(X)$, in the class, $X$ and $\omega$ is the average time it takes to generate an influential test:

\begin{equation}\label{eq:5}
\omega=\frac{t-1}{\left|\tau\left(X,c\right)\right|} 
\end{equation}

Finally, Equation \ref{eq:1} can be rewritten as follows:
\begin{equation}\label{eq:6}
    \begin{aligned}
T\left(X\right) = & \frac{1}{|Criteria|}\sum_{c\in Criteria}{c^{level}(X)} \\
\times & \frac{1}{\left(1+\omega\right)^{\left\lceil\frac{|\tau\left(X,c\right)|}{NOM(X)}\right\rceil-1}}  
    \end{aligned}
\end{equation}

The testability, $\bar{T}(M)$, of a component, $M$, including $n$ classes, can be computed as the average testability of its classes:
\begin{equation}\label{eq:7}
\bar{T}\left(M\right)\ =\ \frac{1}{n}\sum_{i=1}^{n}{T_i(X)}
\end{equation}

The mathematical model has been used to label the samples, representing class components, with their testability measures at run time. The following section describes how to use the labeled samples to learn a testability prediction model.

\subsection{Testability prediction}\label{sec:testability-measurement-framework}
\begin{figure}
    \centering
    \includegraphics[width=0.90\linewidth]{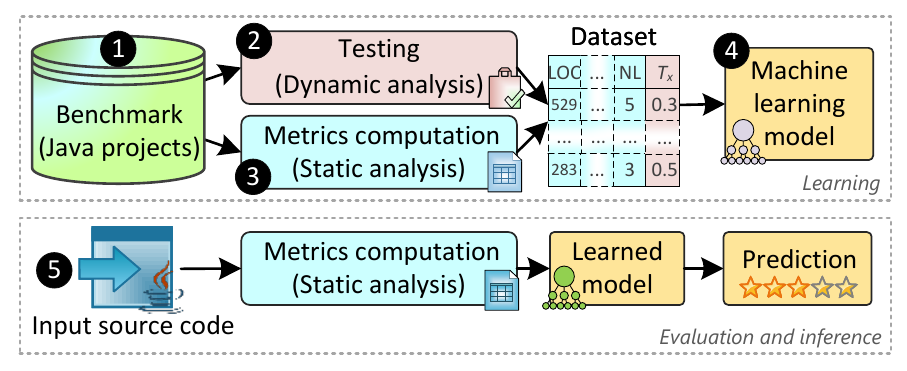}
    \caption{Testability prediction process.}
    \label{fig:framework}
\end{figure}

\noindent 
The proposed testability prediction model aims to estimate the value computed by Equation \ref{eq:6} based on the source code metrics. This way, there will be no need to generate test data and run the class under test. 
The learning algorithm finds the parameters of a real-valued function $f$ that maps the vector of the source code metrics, $\vec{v}$, of a given class, $X$, to its testability value, $T(X)$, \textit{i.e.}, $f: \vec{v} \in \mathbb{R}^n\rightarrow T\in\mathbb{R}$, with minimum possible error. 
Figure \ref{fig:framework} illustrates the testability prediction process, which consists of two learning and inference phases. A detailed description of each phase is as follows:

\begin{enumerate}[label=\arabic*.]
\item{\textbf{Data collection.} 
    Java classes from various software systems are collected to be used as benchmarks for testability prediction.
}
\item{\textbf{Target value computation.}
    Each class in the benchmark is tested to obtain dynamic metrics used by testability mathematical model.
    The target value for the regression models is then computed using the testability mathematical model (Equation \ref{eq:6}).
}
\item{\textbf{Feature vector construction.} 
    Each class in the prepared dataset is converted to a feature vector, $\vec{v}$, in which each feature indicates a source code metric.
}
\item{\textbf{Model training.}
     An ensemble of multilayer perceptron \cite{Goodfellow2016}, random-forest \cite{Breiman2001}, and histogram-based gradient boosting regressors \cite{NIPS2017, Guryanov2019} is trained on the dataset.
    The training samples consist of a vector of source code metrics labeled by the testability of a class within the dataset. Each of the base regressors is trained using a five-fold cross-validation method.  
}
\item{\textbf{Model inference.}
    The learned model is used to predict the testability of a given class based on the static metrics.
} 
\end{enumerate}

A detailed description of the abovementioned steps is given in the following sections. The learning process begins with computing source code and runtime metrics used as independent and dependent variables.

\subsection{Metrics computation}\label{sec:metrics-computation}
\noindent
Two sets of dynamic and static metrics, respectively evaluated at runtime and compile-time, are used to construct the testability prediction model.
The dynamic metrics are computed while running the program under test to generate test data. Section \ref{sec:run-time-information} describes the dynamic metrics as coverage metrics and the approach used to compute them. In contrast to dynamic metrics, static metrics are computed without any need to execute the program under test. The static metrics are the subject of Section \ref{sec:compile-time-information}.

\subsubsection{Dynamic metrics}\label{sec:run-time-information}
\noindent
The runtime metrics include branch coverage, statement coverage, and the size of the minimized test suite provided by EvoSuite \cite{Arcuri2016}. These runtime metrics were used as the parameters in Equation \ref{eq:6} to evaluate each class's testability in the prepared dataset.

 EvoSuite is an evolutionary test data generator tool that uses the whole test suite generation technique \cite{Fraser2013}. A candidate solution (chromosome) in the whole test suite generation technique is a test suite consisting of a variable number of test cases. A population of candidate solutions is evolved during the evolutionary search using operators imitating natural evolution such as crossover and mutation. Individuals are selected for reproduction based on their fitness, i.e., an estimation of how close they are to the optimal solution. 
 The evolutionary process terminates once the test budget is exhausted or a hundred percent coverage is achieved. The resulting test suite is finally minimized to include only influential test cases affecting the test suite's coverage. 
 
 EvoSuite generates JUnit test classes for each class in a given Java project. It saves the test suite size and its coverage information for all the classes of the Java project in a CSV file. The authors have shown that their whole test suite generation approach outperforms the other test data generation approaches \cite{Fraser2013, Fraser2014, Panichella2020}. 
 EvoSuite \cite{Arcuri2016} uses a stochastic approach. Each time it generates tests for a class, those tests may differ in number, length, construction, and attained coverage. 
 Therefore, EvoSuite was run multiple times on each project to generate test suites with different random seeds, and then the results were averaged.

\subsubsection{Static metrics}\label{sec:compile-time-information}
\noindent
The compile-time metrics include 262 metrics computed using the compiler front-end analysis.  These metrics are used to convert a Java class into a feature vector. 
The vectorized representation of the source code is improved as the number of metrics increases. Initially, 40 well-known source code metrics listed in Table \ref{tabel:metrics} were selected.
The 'CS' abbreviation refers to class-level metrics, and the 'PK' abbreviation refers to package-level metrics. 
Afterward, these features were extended by applying a systematic metric generation technique to seven metrics evaluated for the methods of a class (CS) and classes within each package (PK). Techniques applied to enhance the feature space are described in Section \ref{sec:systematic-metrics}.

In addition, this paper defines and computes a new set of metrics, called \emph{lexical} metrics, to represent the lexical properties, such as the number of identifiers, operators, and imports used in a source code file. 
Lexical metrics are listed in Table \ref{tabel:lexical-metrics}.
A Java source file may contain several classes as well as a class may be enclosed by another class (nested classes). The proposed lexical metrics capture the size and complexity of the Java source file containing the class under test. Since these metrics are computed for each Java file separately, we refer to these metrics as file-level lexical metrics.

Most of the metrics listed in Table \ref{tabel:metrics} have already been proposed in the literature, including Chidamber and Kemerer (C\&K) metrics \cite{Chidamber1994}, HS metrics \cite{Henderson-Sellers1995}, MOOD metrics \cite{Harrison1998}, QMOOD metrics \cite{Bansiya2002}, MTMOOD metrics \cite{Khan2009} and Custom metrics introduced in \cite{ArcelliFontana2016}.
However, only a few of these metrics are utilized to measure the testability of the software \cite{NUNEZVARELA2017164}. 
Each source code metric seemingly affecting the testability is mentioned in Tables \ref{tabel:metrics} and  \ref{tabel:lexical-metrics}. In addition, all the metrics, such as the number of blank lines, number of comment lines, and comment-to-code ratio that could not affect the testability, were not added to the features vector.

\begin{table}
    \centering
     \caption{Source code metrics.}
     \label{tabel:metrics}
    \resizebox{0.90\linewidth}{!}{%
        \begin{tabular}{llllll} 
            \hline
            Subject                       & Metric                                                                                                                                                      & Full name                                                                                                                                                                                                                                                                                                                                                                           & CS                                                                     & PK                                                                       & Sum  \\ 
            \hline
            Size& \begin{tabular}[c]{@{}l@{}}\textbf{LOC }\\\textbf{NOST}\\NOSM\\NOSA\\NOIM \\NOIA\\NOM\\NOMNAMM \\NOCON\\\textbf{NOP}\\NOCS\\NOFL \end{tabular}              & \begin{tabular}[c]{@{}l@{}} Line of code\\Number of (NO.) statements\\NO. static methods\\NO. static attributes\\NO. instance methods\\NO. instance attributes\\NO. methods\\NO. not accessor or mutator methods\\ NO. constructors\\NO. parameters\\NO. classes\\NO. files \end{tabular}                                      & \begin{tabular}[c]{@{}l@{}} 36\\36\\1\\1\\1\\1\\1\\1\\1\\10\\0\\0\end{tabular} & \begin{tabular}[c]{@{}l@{}} 15\\15\\1\\1\\1\\1\\1\\1\\1\\0\\1\\1\end{tabular} & 128\\ 
            \hline
            Complexity                    & \begin{tabular}[c]{@{}l@{}}\textbf{CC}\\\textbf{NESTING }\\\textbf{PATH}\\\textbf{KNOTS }\end{tabular}                                                     & \begin{tabular}[c]{@{}l@{}}Cyclomatic complexity\\Nesting block level\\NO. unique paths \\NO. overlapping jumps \end{tabular}                                                                                                                                                                                                                         & \begin{tabular}[c]{@{}l@{}} 40\\4\\10\\10 \end{tabular}                        & \begin{tabular}[c]{@{}l@{}} 20\\4\\0\\0 \end{tabular}                                   & 88\\ 
            \hline
           \begin{tabular}[c]{@{}l@{}}Cohesion\\~~~~and \\Coupling \end{tabular} & \begin{tabular}[c]{@{}l@{}}NOMCALL\\DAC\\ATFD\\LOCM\\CBO\\RFC\\FANIN\\FANOUT \\DEPENDS\\DEPENDSBY\\CFNAMM \\\end{tabular} & \begin{tabular}[c]{@{}l@{}} NO. method calls\\Data abstraction coupling\\Access to foreign data\\Lack of cohesion in methods\\Coupling between objects\\Response set for a class\\NO. incoming invocations\\NO. outgoing invocations\\All dependencies of class\\Entities depended on class\\Called foreign not accessors or mutators \end{tabular} & \begin{tabular}[c]{@{}l@{}} 1\\1\\1\\1\\1\\1\\1\\1\\1\\1 \\1 \\\end{tabular}       & \begin{tabular}[c]{@{}l@{}} 0\\0\\0\\0\\0\\0\\0\\0\\0\\0 \\0 \\\end{tabular}                & 11\\ 
            \hline
            Visibility                    & \begin{tabular}[c]{@{}l@{}} NODM \\NOPM \\NOPRM\\NOPLM \\NOAMM \end{tabular}                                                                                & \begin{tabular}[c]{@{}l@{}} NO. default methods\\NO. private methods\\NO. protected methods\\NO. public methods\\NO. accessor methods \end{tabular}                                                                                                                                                                                                   & \begin{tabular}[c]{@{}l@{}} 1\\1\\1\\1\\1\end{tabular}                         & \begin{tabular}[c]{@{}l@{}} 1\\1\\1\\1\\1\end{tabular}                                  & 10\\ 
            \hline
            Inheritance                   & \begin{tabular}[c]{@{}l@{}} DIT\\NOC\\NOP\\NIM\\NMO\\NOII\\NOI\\NOAC \end{tabular}                                                                          & \begin{tabular}[c]{@{}l@{}} Depth of inheritance tree\\NO. children\\NO. parents\\NO. inherited methods\\NO. methods overridden\\NO. implemented interfaces\\NO. interfaces\\NO. abstract classes \end{tabular}                                                                                                                           & \begin{tabular}[c]{@{}l@{}} 1\\1\\1\\1\\1\\1\\0\\0\end{tabular}                & \begin{tabular}[c]{@{}l@{}} 0\\0\\0\\0\\0\\0\\1\\1\end{tabular}                         & 8\\ 
            \cline{3-6}
             \rowcolor[rgb]{0.985,0.985,0.985} Sum  & 
            ~ &
            40 & 
            175  & 
            70 &
             245\\
            \hline
        \end{tabular}
    }
\end{table}

\begin{table}
    \centering
    \caption{Lexical metrics computed for each source code file.}
    \label{tabel:lexical-metrics}
    \resizebox{0.55\linewidth}{!}{%
        \begin{tabular}{ll}
            \hline
            Metric  & Full name   \\ \hline
            NOTK    & Number of (NO.) tokens  \\
            NOTKU   & NO. unique tokens                 \\
            NOID    & NO. identifiers                   \\
            NOIDU   &  NO. unique identifiers            \\
            NOKW    &  NO. keywords                      \\
            NOKWU   &  NO. unique keywords               \\
            NOASS   &  NO. assignments                   \\
            NOOP    &  NO. operators without assignments \\
            NOOPU   &  NO. unique operators  \\
            NOSC    &  NO. semicolons \\
            NODOT   &  NO. dots  \\
            NOREPR  &  NO. return and print statements   \\
            NOCJST  &  NO. conditional jumps  \\
            NOCUJST &  NO. unconditional jumps           \\
            NOEXST  & NO. exceptions                    \\
            NONEW   & NO. \texttt{new} objects instantiation     \\
            NOSUPER &  NO. \texttt{super} calls   \\ \hline
        \end{tabular}%
    }
\end{table}

\subsubsection{Systematic metrics}\label{sec:systematic-metrics}
\begin{figure}[!t]
    \centering
    \includegraphics[width=0.75\linewidth]{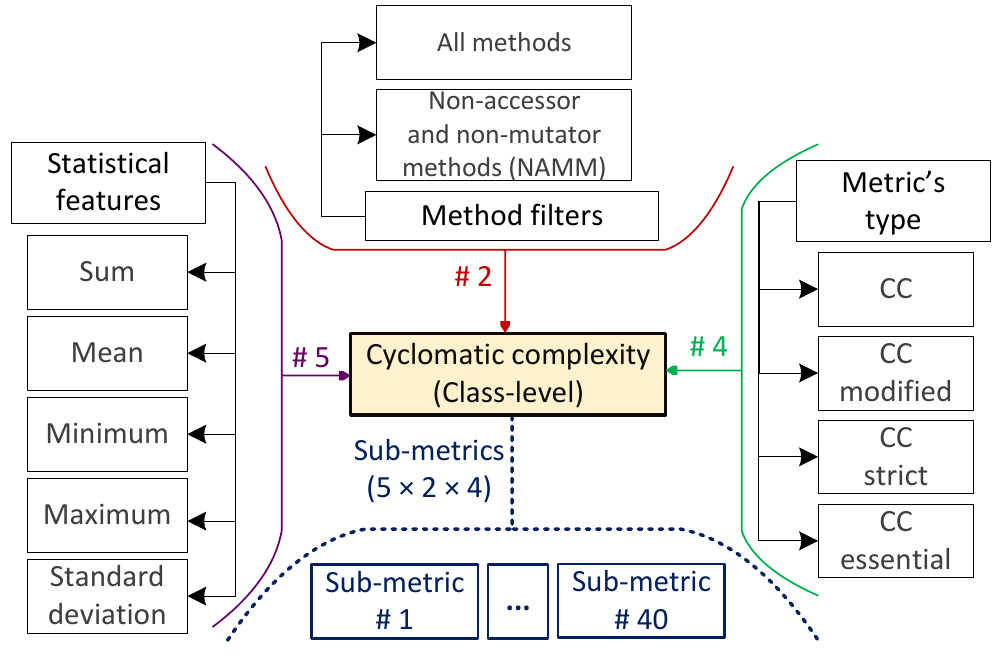}
     \caption{Sub-metrics derived from cyclomatic complexity.}
     \label{fig:cyclomatic-complexity-submetrics}
\end{figure}

\noindent
This paper offers a \emph{systematic} approach to enhancing the feature space by deriving \emph{sub-metrics} from method-level metrics. By selecting seven method-level metrics, bolded in Table \ref{tabel:metrics}, 200 sub-metrics could be derived. 
Sub-metric is a new concept introduced in this paper. A sub-metric is a metric derived from an existing software metric. For instance, cyclomatic complexity is a known metric defined and computed for the program's methods. Average, sum, min, max, and standard deviation are the five operators applied to the cyclomatic complexities of a class method to compute different cyclomatic complexity sub-metrics for the class. In addition, as shown in Figure \ref{fig:cyclomatic-complexity-submetrics}, there are four different computations for cyclomatic complexity \cite{SciTools2020} to each of which these operators are applied. These operators are applied once with and another time without considering the \emph{accessor} and \emph{mutator}  methods \cite{ArcelliFontana2016}. In total, by applying this set of operators, 40 effective sub-metrics have been generated for cyclic complexity. 
According to Figure \ref{fig:cyclomatic-complexity-submetrics}, sub-metrics derived from cyclomatic complexity are as follows: 

\begin{itemize}[noitemsep, labelwidth=!, labelindent=0pt, topsep=0pt]
    \item{Sub-metric \#1: CC\_Sum\_All\_Methods} 
    \item{Sub-metric \#2: CC\_Sum\_NAMM}
    \item{Sub-metric \#3: CC\_Mean\_All\_Methods}
    \item{Submetrics \#4: CC\_Mean\_NAMM} 
    \item{. . .}
     \item{Submetric \#39: CCEssential\_SD\_All\_Methods} 
    \item{Submetric \#40: CCEssential\_SD\_NAMM} 
\end{itemize}

\begin{figure}[!h]
    \centering
    \includegraphics[width=0.85\linewidth]{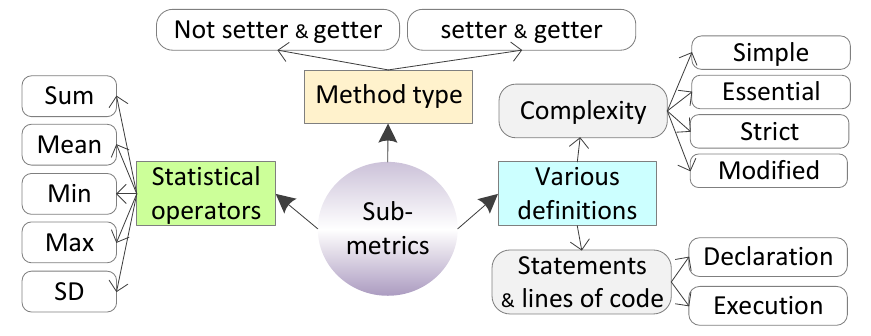}
    \caption{Sub-metric construction map for a given class.}
    \label{fig:systematic-metric}
\end{figure}

These operators were applied to six other method-level metrics. The same process has been applied to derive package-level metrics from class-level.
A class's interactions with the other classes, or in other words, the context of a class, affects its functionality. That is why statistical operators were applied at the package level to all the classes in the enclosing package. 
The metrics computed for a package were attached to the feature vectors of all its including classes. In summary, sub-metrics for a class can be derived from the method-level metrics using four different viewpoints, shown in Figure \ref{fig:systematic-metric}.
The number of sub-metrics corresponding to each primary metric is shown in Table \ref{tabel:metrics}. 

One can observe that some systematically constructed metrics have already been defined and used in literature, \textit{e.g.}, WMC \cite{Chidamber1994}. 
Duplicate metrics were eliminated from the final set. Constructing sub-metrics makes it possible to study the impact of each high-level metric in more detail. For example, one can answer questions such as whether or not a class with a high complexity variance is testable. 

A similar operation can be performed to create additional package-level metrics from class-level metrics. 
In total, 57 metrics were computed, and seven were extended systematically, resulting in 262 different metrics.
The complete list of metrics and samples used in the experiments is available on the testability prediction dataset published at \href{https://doi.org/10.5281/zenodo.4650228}{\textit{https://doi.org/10.5281/zenodo.4650228}}.

\subsection{Dataset preparation}\label{sec:data-preparation}
\noindent 
This section describes dataset schema and preprocessing operations before data is fed to machine learning models. 
The format of data used in the dataset is described in Section \ref{sec:data-representation}. Before using the collected data for learning, the data should be cleaned and normalized. Section \ref{sec:data-preprocessing} describes the dataset preprocessing as a known machine learning stage.

\subsubsection{Data representation}\label{sec:data-representation}
\begin{figure}[!h]
    \centering
    \includegraphics[width=0.90\linewidth]{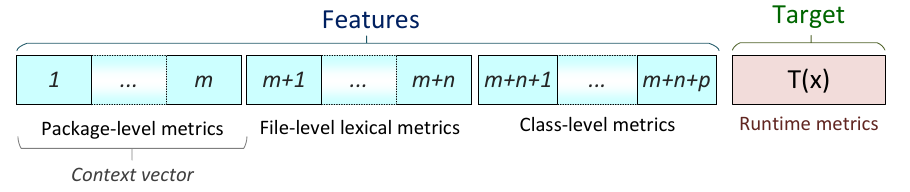}
    \caption{Structure of each sample in the testability prediction dataset.}
    \label{fig:dataset-schema}
\end{figure}

\noindent 
The prepared testability prediction dataset is in a tabular format. Each row represents a class instance, and each column represents a metric as an attribute of that class. 
The last column is a numerical variable that expresses the testability value of the class under test, computed by Equation \ref{eq:6}. 
Figure \ref{fig:dataset-schema} illustrates the structure of the source code and runtime metrics in the testability prediction dataset.
The context vector consists of the package-level metrics, repeated for all classes in a package. These metrics are specified in the 5\textsuperscript{th} column of Table \ref{tabel:metrics}. 
Adding package-level metrics to the feature vector of each class in the package facilitates exploiting the interactions among the features while classifying the class elements. 
The labeled vectors should be preprocessed before they can be fed to the learning algorithms.

\subsubsection{Data preprocessing}\label{sec:data-preprocessing}
\begin{figure}
    \centering
    \includegraphics[width=0.90\linewidth]{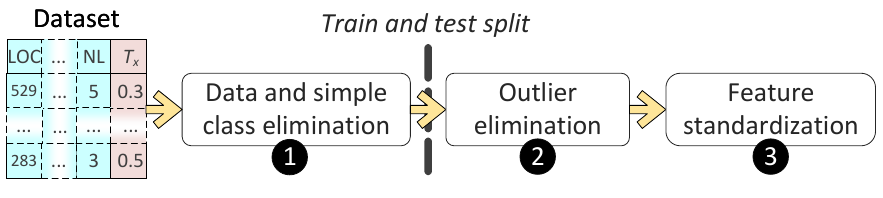}
    \caption{Preprocessing steps.}
    \label{fig:preprocessing-pipeline}
\end{figure}

\noindent
Data must be cleaned before building any machine learning model to ensure all the samples' pertinency and authenticity and avoid incorrect training. The data preprocessing steps are illustrated in Figure \ref{fig:preprocessing-pipeline}. The dataset is prepared to be used for building prediction models in three steps:

First, data classes (classes that only contain data fields, mutators, and accessor methods) and simple classes (classes with a LOC less than 5) are removed. The reason is that data and simple classes are inherently testable and do not impose a high test effort.

Second, data samples for which one or more metrics are very high or very low are identified and removed  as outliers by applying the local outlier factor (LOF) algorithm \cite{Breunig2000}. These samples negatively affect parameter tuning during the learning process.

Third, metrics are standardized by scaling their values into the same range \cite{Scikit-learn2020}. Most learning algorithms, such as artificial neural networks, are susceptible to the range of independent variables and simply biased towards the most significant values.

It is important to note that steps two and three are performed after partitioning data into train and test sets. This way,  the methodological mistake related to leaking information from train to test data is avoided. 

 After the preprocessing stage, the dataset gets ready to learn the testability prediction model described in Section \ref{sec:prediction-models}.

\subsection{Prediction models}\label{sec:prediction-models}
\noindent 
Since the testability, $T(X)$, of a class, $X$, is continuous in the interval $[0, 1]$, regression techniques are applied to construct a machine learning model for predicting testability. 
The learning algorithms, described in Section \ref{sec:learning-algorithms}, constitute an ensembled meta-estimator. The testability measures provided by the learned meta-estimator model may require further modifications, described in Section \ref{sec:inference-algorithms}.

\subsubsection{Learning algorithms}\label{sec:learning-algorithms}
\noindent 
This paper examines six regression models from different families of learning algorithms to find the best model for testability prediction. Firstly, a linear regression model \cite{Zhang2004} is trained and evaluated. 
Five off-the-shelf regressors are also built to capture any possible non-linear relationships between source code metrics and testability. 
The non-linear regressors are: support vector machine regressor (SVMR) \cite{Chih-Chung2011}, decision tree regressor (DTR) \cite{Scikit-learn2020}, random forest regressor (RFR) \cite{Breiman2001}, histogram-based gradient boosting regressor (HGBR) \cite{NIPS2017, Guryanov2019}, and multi-layer perceptron regressor (MLPR) \cite{Goodfellow2016}.
This paper uses the histogram-based gradient boosting algorithm instead of naïve gradient boosting. A significant problem of naïve gradient boosting is that it is slow to train the model, particularly when using the model on large datasets containing ten thousand samples with hundreds of continuous features. The histogram-based gradient boosting algorithm discretizes the continuous input variables to a few hundred unique values or bins (typically 256 bins), which tremendously reduces the number of splitting points to consider and increases the learning speed. 
Finally, a voting regressor (VoR) \cite{Scikit-learn2020} is used to compute the weighted average of predictions made by the regressors as the ultimate testability measure. 
A voting regressor is an ensemble meta-estimator that fits several base regressors, each on the whole dataset. Then it averages the individual predictions of the base regressors to form a final prediction \cite{Scikit-learn2020}.

Before training the regressors, the relevant hyper-parameters should be configured. A grid search strategy with cross-validation \cite{Bengio2012} finds the optimal hyperparameters for each model during the training process. In this way, the most appropriate configuration for the hyperparameters is made.
 
\subsubsection{Estimation  algorithm}\label{sec:inference-algorithms}
\noindent 
Testability estimation algorithm, shown in Algorithm \ref{alg:inference-testability}. It receives a class under test and its enclosing project along with a testability learned model, its hyperparameters, and the learning dataset. The algorithm computes and returns the testability of the class under test in a given project as output. 
By default, a simple or data class’s testability equals 1. Otherwise, the model is asked to estimate the testability by calling the predict method of the learned model. 

The proposed regression model is assumed to compute the class testability in the interval of [0, 1]. However, if the input (independent variables) falls far from the learned distribution, the model may compute the testability as a value out of the specified interval. In such rare cases, the algorithm changes the computed testability value to 0 or 1 depending on whether it is less than zero or greater than one.

\begin{center}
    \begin{minipage}{0.95\linewidth}
        \vspace{-1mm}
        \small
        \begin{algorithm}[H]
            \small
            \onehalfspacing
            \caption{EstimateTestability}\label{alg:inference-testability}
            \DontPrintSemicolon
            \setcounter{AlgoLine}{0}
            \LinesNumbered
            \SetKwFunction{computeMetrics}{computeMetrics}
            \SetKwFunction{TPM}{TPM}
            \SetKwFunction{getMetricsUsed}{getMetricsUsed}
            \SetKwFunction{normlizeMetrics}{normlizeMetrics}
            \SetKwFunction{RandInt}{RandInt}
            \SetKwFunction{predict}{predict}
            \SetKwInput{KwData}{Input}
            \SetKwInput{KwResult}{Output}
            \KwData{
                1. ClassUncerTest (CUT): The class whose testability is going to be predicted,\\
                2. TestabilityPredictionModel (TPM): The learned testability prediction model,\\
                3. Dataset (DS): The labeled dataset used to train the testability prediction model,\\
                4. Hyprparameters ($\theta$): The set of parameters used to configure the regression model,\\
                5. ProjectUnderTest (PUT): Project including the class under test.
            }
            \KwResult{Testability}
            \BlankLine
            
            metricsNames $\gets$ \TPM(DS, $\theta$).\getMetricsUsed(); 
            
            classMetricsVector $\gets$ \computeMetrics(CUT, metricsNames, PUT);
            
            isSimpleClass $\gets$ classMetricsVector[”CSLOC”] $<$ 5;
            
            isDataClass $\gets$ classMetricsVector[”CSNOMNAMM”] $==$ 0 $\land $
            (classMetricsVector[”CSNOIA”] $+$ classMetricsVector[”CSNOSA”]) $>$ 0;
            
            \tcc{{\color{Green}Check whether ClassUnderTest is a simple or data class}}
            \eIf{isSimpleClass $\lor$ isDataClass} {
                
                Testability $\gets$ 1;
            
        }{
        
      normalizedMetricsVector $\gets$ {[} {]}; 

        \ForEach{metric \textbf{in} classMetricsVector }{
             \tcc{{\color{Green}Normalize metrics with respect to train set}}
             metric $\gets$ normlizeMetric(DS, metric.name, metric.value); 
             
             normalizedMetricsVector.add(metric);
        
    }
    Testability $\gets$ \TPM(DS, $\theta$).\predict(normalizedMetricsVector); 

            \uIf{Testability $<$ 0} {
                Testability $\gets$ 0;
            }
         \ElseIf{Testability $>$ 1} {
             Testability $\gets$ 1;
        }
    }
  
    \textbf{return} Testability;

        \end{algorithm}
    \end{minipage}
\end{center}

The computational complexity of Algorithm \ref{alg:inference-testability} depends on two main steps in lines 2 and 13, which are responsible for calculating source code metrics and predicting testability.
Line 2 computes the source code metrics to convert a class in the project under test into a feature vector. The calculation of each source code metric needs a depth-first traversal of the program parse tree. The order of the depth-first traversal is $ O(n) $, where $ n $ is the number of nodes of the parse tree \cite{Cormen2022}. Constructing a feature vector with length $ d $ in this way is of order $ O(d \cdot n) $. The algorithm's time complexity to build the parse tree is $ O(k) $, where $ k $ is the number of tokens in the project under test \cite{Aho2006}. Therefore, the order of the function to compute a features vector is $ O(k + d \cdot n) $. The number of nodes, $ n $, of the program's parse tree is more than the number of tokens, $ k $. As a result, the final order is $ O(d \cdot n) $ , which is equal to $ O(n) $ as the length of the feature vector is fixed.

Line 13 predicts testability based on the vector of source code metrics using the proposed testability prediction model. The ensemble meta-estimator model is based on the RFR, HGBR, and MLPR models. The prediction of a given sample with PFR and HGBR models is of the order $ O(t \cdot \log d) $, where $ t $ is the number of trees and $ d $ is the number of features assuming that the trees are free to grow to the maximum height, $ O(\log d) $ \cite{Xavier2014}. The computational complexity of predicting a sample with the MLPR model depends on the size of the features vector, the number of hidden layers, and the size of each layer. Both the space and time complexity of a multi-layer perceptron network is $ O(H \cdot (K + d)) $, where $d$ is the input dimension, $ H $ is the number of hidden units, and $ K $ is the number of outputs \cite{Alpaydin2020}. In the trained model, the size of the features vector, the number of hidden layers, and the size of each layer are fixed values. Moreover, the only output of the regressor model is testability. Therefore, the computational complexity of the proposed MLPR model's feed-forward pass used in prediction time is linear in terms of the input dimension, $ d $. 
Overall, the computational complexity of the proposed testability prediction algorithm for a given class is of the order $ O\left(d \cdot n\ +2 \cdot t \cdot \log d + H \cdot \left(1+d\right) \right) $. Since the parameters $ d $, $ t $, and $ H $ are fixed numbers, the complexity of the testability prediction algorithm concerning the program's parse tree size, $  n $, is $ O(n) $ in the worst case. It concludes that the computational complexity of the proposed algorithm is linear in terms of the input program size.

\subsection{Implemented tool}
\noindent 
The proposed testability prediction approach has been implemented in Python 3 and can be used as a standalone tool. Both the implemented tool and the evaluation dataset are available at a public GitHub repository, \textit{\href{https://github.com/m-zakeri/ADAFEST}{https://github.com/m-zakeri/ADAFEST}}. 
Complete documentation of the source code and the dataset are also available on \textit{\href{https://m-zakeri.github.io/ADAFEST}{https://m-zakeri.github.io/ADAFEST}}. 

Code coverage information for each Java class was extracted from EvoSuite reports using a python script. Source code metrics for each class were extracted using SciTools Understand API \cite{SciTools2020}. 
The Understand software kit provides a command-line tool, \emph{und}, that analyzes project files and creates a database containing code entities and their relationships. It offers an API to query the database and compute the source code metrics. The proposed tool first extracts the required metrics for a given class with Understand API and then calls the machine learning module.
Data preprocessing and machine learning algorithms were implemented using the Scikit-learn \cite{Scikit-learn2020}, an open-source python data analysis framework. 

\section{Experiments and evaluations}\label{sec:experiments}
\noindent 
In this section, the experiments with the proposed methodology and thier results are reported to answer the following research questions:

\begin{itemize}
    \item{\textbf{RQ\textsubscript{1}} 
       \textit{What is the best machine learning model to predict testability? }
    }
     \item{\textbf{RQ\textsubscript{2}}
         \textit{Does the newly introduced sub-metrics, lexical metrics, and package-level metrics improve the testability prediction model accuracy by enhancing the feature space?}
    }
      \item{\textbf{RQ\textsubscript{3}}
         \textit{Which source code metrics affect the testability of a class more than others?}
        }
      \item{\textbf{RQ\textsubscript{4}}
          \textit{Is it possible to improve software testability by improving influential source code metrics via automatic refactoring?}
        }
     \item{\textbf{RQ\textsubscript{5}}
      \textit{Does refactoring for testability improve other quality attributes?}
     }
\end{itemize}

\subsection{Experimentation setup}
\noindent 
All experiments were performed on Windows 10 (x64) machine with a 2.6 GHz Intel\textregistered~ Core\texttrademark~i7 6700HQ CPU and 16 GB RAM. The static metrics extraction and preprocessing of the primary dataset on this machine took about 25 hours. 
The SF110 corpus \cite{Fraser2014}, containing 23,886 classes from 110 different Java projects, was used as the benchmark to create train and test sets for the machine learning models.
EvoSuite \cite{Arcuri2016} was run \emph{five} times on each project with different random seeds and averaged the results to minimize the randomness effects caused by evolutionary test data generation.
A general timeout of \emph{six} minutes per class was given to each execution to ensure that the experiments with Java classes in SF110 finished within a predictable time. 
The experimental results showed that when repeating the test data generation five times, the mean standard deviation of the coverage provided by the generated test data reduces to 0.024, which is acceptable.

Seven regression models, described in Section \ref{sec:learning-algorithms}, were trained and compared to find the best machine learning algorithm for predicting testability.  
In addition to the original dataset (DS1), different subsets of source code metrics were used to construct four datasets (DS2 to DS5). 
The aim is to experiment with the impact of context vector (package-level) metrics, lexical metrics, and systematically generated metrics on the effectiveness of predictive models.
 
Table \ref{tabel:datasets} represents the datasets and their schema. 
The DS1 dataset was initially taken from the SF110 \cite{Fraser2014} benchmark. SF110 contains 110 Java projects with more than 23,800 classes, which is big enough for the learning tasks.
The preprocessing steps applied to DS1 led to the elimination of 4,100 classes, and the final dataset includes 19,750 samples. 
DS2 to DS5 datasets were built by removing specific columns, shown in Table \ref{tabel:datasets}, from DS1. Each model was trained and evaluated on all datasets in Table \ref{tabel:datasets}. 
Datasets were randomly split to train and test sets before applying the model selection process.  
Models were trained on 70\% of the data (14,000 Java classes) and tested on the remaining 30\% (5,750 samples).
 
\begin{table}
    \centering
    \caption{Datasets used in the experiments.}
    \label{tabel:datasets}
    \resizebox{0.85\linewidth}{!}{%
        \begin{tabular}{lll}
            \hline
            Dataset                                                 & Applied preprocessing                                                                                                                             & \# Metrics \\ \hline
            \begin{tabular}[c]{@{}l@{}}
                DS1 \end{tabular} & \begin{tabular}[c]{@{}l@{}}Simple and data classes elimination, \\ outliers elimination, and metric standardization\end{tabular} & 262        \\
             \rowcolor[rgb]{0.985,0.985,0.985} DS2                                                     & DS1 + Feature selection                                                                                                                           & 20         \\
            DS3                                                     & DS1 + Context vector elimination                                                                                                                  & 194        \\
             \rowcolor[rgb]{0.985,0.985,0.985} DS4                                                     & DS1 + Context vector \& lexical metrics elimination                                                                                              & 177        \\
            DS5                                                     & DS1 + Systematically generated metrics elimination                                                                                                & 71         \\ \hline
        \end{tabular}%
    }
\end{table}

\subsection{Prediction models evaluation}
\noindent
Concerning RQ\textsubscript{1}, the effectiveness of each learned model was measured with standard metrics used for evaluating the performance of regression models, including mean absolute error (MAE), mean square error (MSE), root mean square error (RMSE), median absolute error (MdAE), and R\textsuperscript{2} score. 
Table \ref{tabel:rq1} shows the evaluation metrics for all models learned using DS1. The last row shows the result of the VoR model, which is an ensemble of RFR, HGBR, and MLPR models. These three models were selected because of their relatively higher R\textsuperscript{2} scores in the model selection process. 
In Table \ref{tabel:rq1}, the best value obtained for each evaluation metric is bolded. According to the evaluation results given in this table, it concluded that:

\begin{itemize}
    \item{The ensemble meta-regressor, VoR, performs better than the individual model in terms of MSE, RMSE, and R\textsuperscript{2} score, revealing that combining different models may result in more accurate predictions. 
        
    }
    \item{Overall, non-linear models denote relatively better performance which confirms non-linear relationships between source code metrics and testability.
    }
\end{itemize}

 \begin{table}
     \centering
     \caption{Performance of different testability prediction models on DS1.}
     \label{tabel:rq1}
     \resizebox{0.75\linewidth}{!}{%
     \begin{tabular}{llllll} 
         \hline
         \rowcolor[rgb]{1, 1, 1} Model      & MAE                & MSE                 & RMSE               & MdAE               & R\textsuperscript{2}-score            \\ 
         \hline
         Linear                                 & 0.14745            & 0.03748             & 0.19361            & 0.11405            & 0.56899             \\
         SVMR                                   & 0.15865            & 0.04423             & 0.21032            & 0.11696            & 0.47144             \\
         DTR                                    & 0.14620            & 0.04090             & 0.20223            & 0.10485            & 0.52974             \\
         \rowcolor[rgb]{0.985,0.985,0.985} RFR  & 0.12343            & 0.03013             & 0.17358            & 0.08703            & 0.65354             \\
         \rowcolor[rgb]{0.985,0.985,0.985} HGBR & \textbf{0.11912} & 0.02842             & 0.16859            & \textbf{0.08182} & 0.67319             \\
         \rowcolor[rgb]{0.985,0.985,0.985} MLPR & 0.13676            & 0.03416             & 0.18482            & 0.09868            & 0.60723             \\
         \rowcolor[rgb]{0.920,0.920,0.950} VoR  & 0.11921            & \textbf{0.02801} & \textbf{0.16738} & 0.08231           & \textbf{0.67787}  \\
         \hline
     \end{tabular}
 }
 \end{table}

Table \ref{table:hyperparameters} shows the hyperparameters of each regression model and their best values found for DS1 in the model selection process. RMSE has been used as a metric for scoring and ranking the model's performance along with five cross-fold validation during model selection. For VoR, the only available option is weights used to weight the occurrences of predicted values before averaging. 
The weight array elements in Table \ref{table:hyperparameters} correspond to the Linear, SVMR, DTR, HGBR, RFR, and MLPR models, respectively. 
It is observed that the RFR, HGBR, and MLPR models contribute to the VoR meta estimator.

 \begin{table}[!h]
     \centering
     \caption{Models configurable parameters and results of hyperparameter tuning.}
     \label{table:hyperparameters}
     \resizebox{1.0\linewidth}{!}{%
         \begin{tabular}{llll} 
             \hline
             Model  & 
             \begin{tabular}[c]{@{}l@{}}
                 Hyper-parameter name \\
                  in Scikit-learn
             \end{tabular}&
              Searching values (python statement) & 
             Best value   \\ 
             \hline
              \rowcolor[rgb]{0.985,0.985,0.985} Linear & \begin{tabular}[c]{@{}l@{}}loss\\penalty\\learning\_rate\\max\_iter\end{tabular}                      & \begin{tabular}[c]{@{}l@{}}{[}'squared\_loss', 'huber']\\{[}'l2', 'l1', 'elasticnet']\\{[}'invscaling', 'optimal', 'constant', 'adaptive']~\\range(50, 1000, 50)\end{tabular} & \begin{tabular}[c]{@{}l@{}}'huber'~\\'l2'~\\'invscaling'~\\50~\end{tabular}           \\ 
             \midrule
             SVMR   & \begin{tabular}[c]{@{}l@{}}kernel~\\nu~\end{tabular}                                                  & \begin{tabular}[c]{@{}l@{}}{[}'linear', 'rbf', 'poly', 'sigmoid']~\\{[}0.25, 0.5, 0.75, 1.0]~\end{tabular}                                                                    & \begin{tabular}[c]{@{}l@{}}'rbf'~\\0.5~\end{tabular}                                  \\ 
             \midrule
              \rowcolor[rgb]{0.985,0.985,0.985} DTR    & \begin{tabular}[c]{@{}l@{}}criterion~\\max\_depth~\\min\_samples\_split~\end{tabular}                 & \begin{tabular}[c]{@{}l@{}}{[}'mse', 'mae']~\\range(3, 50, 5)~\\range(2, 30, 2)~\end{tabular}                                                                                 & \begin{tabular}[c]{@{}l@{}}'mse'~\\8~\\28~\end{tabular}                               \\ 
             \midrule
             RFR    & \begin{tabular}[c]{@{}l@{}}n\_estimators~\\criterion~\\max\_depth~\\min\_samples\_split~\end{tabular} & \begin{tabular}[c]{@{}l@{}}range(50, 200, 50)~\\{[}'mse', 'mae']~\\range(3, 50, 5)~\\range(2, 30, 2)~\end{tabular}                                                            & \begin{tabular}[c]{@{}l@{}}150~\\'mse'~\\28~\\2~\end{tabular}                         \\ 
             \midrule
              \rowcolor[rgb]{0.985,0.985,0.985} HGBR   & \begin{tabular}[c]{@{}l@{}}
                 loss~\\max\_depth~\\min\_samples\_leaf~\\max\_iter~\end{tabular}           & \begin{tabular}[c]{@{}l@{}}{[}'least\_squares', 'least\_absolute\_deviation']~\\range(3, 50, 5)~\\range(5, 50, 10)~\\range(100, 500, 100)~\end{tabular}                       & \begin{tabular}[c]{@{}l@{}}'least\_squares'~\\18~\\15~\\500~\end{tabular}             \\ 
             \midrule
             MLP    & \begin{tabular}[c]{@{}l@{}}hidden\_layer\_sizes~\\activation~\\learning-rate~\\epoch~\end{tabular}    & \begin{tabular}[c]{@{}l@{}}{[}(256, 100), (512, 256, 100)]~~\\{[}'logistic', 'tanh', 'relu']~\\{[}'constant', 'adaptive']~\\range(100, 500, 50)~\end{tabular}                 & \begin{tabular}[c]{@{}l@{}}(512, 256, 100)~\\'tanh'~\\'constant'~\\100~\end{tabular}  \\ 
             \midrule
              \rowcolor[rgb]{0.985,0.985,0.985} VoR   & weights                                                                                              & \begin{tabular}[c]{@{}l@{}}
                 {[}None, [0, 0, 0, 1/3, 1/3, 1/3], 
                 \\ {[}0, 0, 0, 3/6, 2/6, 1/6], {[}0, 0, 0, 2/6, 3/6, 1/6]{]}
             \end{tabular}   
          & [0, 0, 0, 2/6, 3/6, 1/6]      \\
             \hline
         \end{tabular}
     }
 \end{table}

The ensembled model and its base models were applied 100 times to predict the testability of randomly selected subsets of the SF110 dataset \cite{Fraser2014}, and the difference between the evaluation metrics of the learned regressors was computed.
Afterward, the statistical test, independent t-test, was used to determine whether the ensemble meta-estimator, VoR, performed better than base regressors. 
Table \ref{table:statistical-test-model-selection} shows the p-value of the independent t-test on MSE and R\textsuperscript{2}-score of the VoR model and three base regressors. The p-value less than $0.05$ for MSE and R\textsuperscript{2}-score metrics in all tests indicates that the VoR model predictions are significantly more accurate than three base regressors, \textit{i.e.}, RFR, HGBR, and MLPR.

\begin{table}
    \centering
    \caption{Results of statistical test on different regressors' performance.}
    \label{table:statistical-test-model-selection}
    \resizebox{0.55\linewidth}{!}{%
        \begin{tabular}{lll} 
            \hline
            \multirow{2}{*}{Base regressor} & \multicolumn{2}{l}{Independent t-test p-value}                                                                                  \\ 
            \cline{2-3}
            & ~MSE                                                                                              & R\textsuperscript{2}-score  \\ 
            \hline
            VoR v.s. RFR                    & 
            $2.5311\times{10}^{-24}$ &  $1.4048\times{10}^{-22}$                        \\
            VoR v.s. HGBR                   &  $1.6997\times{10}^{-2}$                                                                                              &  $5.5463\times{10}^{-3}$                        \\
            VoR v.s. MLPR                   & $2.3327\times{10}^{-93}$                                                                                               &  $1.2459\times{10}^{-85}$                        \\
            \hline
        \end{tabular}
    }
\end{table}

Some researchers consider coverage metrics such as branch coverage as a basis for predicting test effectiveness \cite{Robinson2017, Grano2019}. 

\begin{figure}
    \centering
    \includegraphics[width=1.0\linewidth]{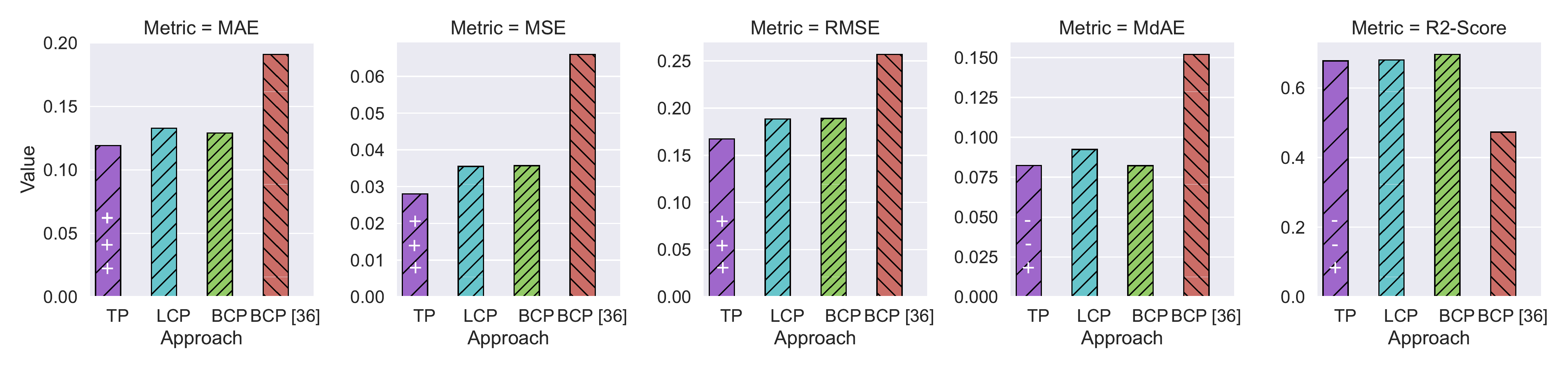}
    \caption{Performance of different testability prediction approaches.}
    \label{fig:testability-approaches}
\end{figure}

The proposed testability prediction (TP) model was compared with the line coverage (LCP), branch coverage (BCP), and branch coverage prediction models by Grano et al. \cite{Grano2019}. 
As shown in Figure \ref{fig:testability-approaches}, the proposed testability prediction model significantly outperformed these three models. 
In Figure \ref{fig:testability-approaches}, the '$ + $' and '$ - $' symbols on the left column of each plot from top to bottom denote the statistical significance between the TP model and the other three models \cite{Grano2019}. The '$ + $' symbol indicates that the TP model outperforms its corresponding model regarding the metric shown by the plot. Similarly, the '$ - $' symbol indicates no statistical difference between the two models concerning the compared metrics shown on the plot. 
As shown in Figure  \ref{fig:testability-approaches}, the prediction error, including MAE, MSE, RMSE, and MdAE, as well as the accuracy, R\textsuperscript{2} score, of the proposed model (TP) is more statistically significant than the other models in almost all (11 out of 15) statistical independent t-tests.
The prediction error of the trained model increases when using statement or line coverage prediction (LCP), branch coverage prediction (BCP), and specific BCP \cite{Grano2019} instead of testability. The underlying reason is that the test effectiveness needs to be normalized with effort. Two classes with the same coverage measure may need different test efforts.

\begin{tcolorbox}[boxrule=0.5pt]
\textbf{RQ\textsubscript{1}}:
\textit{What is the best machine learning model to predict testability?}%

\textbf{Answer to RQ\textsubscript{1}}:
\emph{The experimental results in Table \ref{tabel:rq1} show that the best model is an ensemble of three regressor models, random forest, histogram gradient boosting, and multilayer perceptron. The ensemble model predicts testability with an R\textsuperscript{2}-score of 0.68 and a mean squared error of 0.03. The prediction model is accurate enough to determine testability in practice. The proposed ensemble model has improved mean absolute error by 0.08 (38\%) and R\textsuperscript{2}-score by 0.21 (43\%) due to utilizing new coverage criteria, source code metrics, and learning dataset compared to the relevant machine learning models \cite{Grano2019} for predicting branch coverage.
}
\end{tcolorbox}
    
\subsection{Sub-metrics evaluation}\label{sec:dataset-analysis}
\noindent 
The VoR model was trained on DS2 to DS5 datasets, and then the performance metrics were computed for new models. Table \ref{tabel:rq2} shows the difference between evaluation metrics of the VoR model on DS1 and the other datasets. The last row of the table denotes the mean change in each performance metric. 
The numbers inside parentheses show the p-value of the independent t-test for each metric when testing the prediction models 100 times. 

It is observed that the prediction error of the VoR model increases when switching from DS1 to other datasets. On the other hand, the R\textsuperscript{2} score of models decreases. 
According to resultant p-values, the differences are significant, with a confidence level of $\alpha\ =\ 0.95$ (\textit{p-value} $< 0.05$). The only exception is for the R\textsuperscript{2} score of DS3, in which the difference with DS1 is not statistically significant.
It concludes that all the newly introduced metrics, including source lexical metrics, sub-metrics, and package-level metrics, improve source code vectorization and the testability learning process. 
It is also observed that the impact of lexical metrics is more than other introduced metrics since the performance of the VoR model is highly decreased when using DS4  for training the model.
Besides, automatic feature selection (DS2) can still select the most informative source code metrics while reducing the model complexity.
 
 \begin{table}
     \centering
     \caption{Changes in the VoR performance discarding different code metrics.}
      \label{tabel:rq2}
     \resizebox{0.90\linewidth}{!}{%
         \begin{tabular}{llllll} 
             \hline
             Dataset & MAE         & MSE      & RMSE      & MdAE      & R\textsuperscript{2}-score   \\ 
             \hline
             DS2     & 
             \begin{tabular}[c]{@{}l@{}}
             $ +0.03687 $ \\ ($ 4.2822\times10^{-16} $) 
            \end{tabular}&
            \begin{tabular}[c]{@{}l@{}}
              $ +0.01329 $ \\  ($ 3.1742\times10^{-15} $)
              \end{tabular}&
          \begin{tabular}[c]{@{}l@{}}
               $ +0.03586 $ \\ ($ 2.0290\times10^{-14} $)
               \end{tabular}& 
           \begin{tabular}[c]{@{}l@{}}
              $  +0.04210  $  \\ ($ 1.7748\times10^{-14} $)
              \end{tabular}&
            \begin{tabular}[c]{@{}l@{}}
                $ -0.15280 $   \\ ($ 4.5344\times10^{-13} $)
            \end{tabular} \\
             \rowcolor[rgb]{0.985,0.985,0.985} DS3     & 
             \begin{tabular}[c]{@{}l@{}}
             $ +0.00215 $ \\ ($ 4.8853\times10^{-02} $)
             \end{tabular}&
         \begin{tabular}[c]{@{}l@{}}
             $  +0.00126 $ \\  ($ 1.6745\times10^{-02} $)
             \end{tabular}&
         \begin{tabular}[c]{@{}l@{}}
               $ +0.00373 $  \\($ 1.7160\times10^{-02} $)
               \end{tabular}& 
           \begin{tabular}[c]{@{}l@{}}
              $  +0.00110 $  \\ ($ 3.6002\times10^{-01} $)
              \end{tabular}&
          \begin{tabular}[c]{@{}l@{}} 
              $  -0.01453 $ \\  ($ 6.2538\times10^{-02} $)
              \end{tabular}  \\
             DS4     & 
             \begin{tabular}[c]{@{}l@{}}
             $ +0.18829 $ \\ ($ 5.1738\times10^{-07} $)
             \end{tabular}& 
         \begin{tabular}[c]{@{}l@{}}
             $ +0.20685 $ \\ ($ 3.2255\times10^{-05} $)
             \end{tabular}& 
         \begin{tabular}[c]{@{}l@{}}
             $ +0.31725 $  \\ ($ 2.8377\times10^{-05} $)
             \end{tabular}&
         \begin{tabular}[c]{@{}l@{}} 
            $  +0.11268 $  \\ ($ 6.8180\times10^{-07} $)
        \end{tabular}& 
    \begin{tabular}[c]{@{}l@{}}
             $ -0.32186 $ \\ ($ 4.0390\times10^{-05} $)
             \end{tabular}  \\
             \rowcolor[rgb]{0.985,0.985,0.985} DS5     & 
             \begin{tabular}[c]{@{}l@{}}
             $ +0.00127 $  \\ ($ 4.7907\times10^{-03} $)
             \end{tabular}& 
         \begin{tabular}[c]{@{}l@{}}
             $ +0.00042 $ \\ ($ 2.7829 \times10^{-02} $)
             \end{tabular}& 
         \begin{tabular}[c]{@{}l@{}}
             $ +0.00123 $  \\ $ (2.7210 \times10^{-02} $)
             \end{tabular}&
         \begin{tabular}[c]{@{}l@{}} 
             $ +0.00373 $ \\ $ (1.4899\times10^{-02} $)
             \end{tabular}& 
         \begin{tabular}[c]{@{}l@{}}
             $ -0.00476  $ \\  ($ 2.9298\times10^{-02} $)
             \end{tabular}  \\
        \rowcolor[rgb]{0.920,0.920,0.950}  Average &
         $ 0.0571 $ &
        $  0.0001 $ &
         $ 0.0003 $ &
       $   0.0009 $ &
         $ 0.0012  $
         \\    \hline
         \end{tabular}
     }
 \end{table}

\begin{tcolorbox}[boxrule=0.5pt]
    \textbf{RQ\textsubscript{2}}:
    \textit{Does the newly introduced sub-metrics, lexical metrics, and package-level metrics improve the testability prediction model accuracy by enhancing the feature space?}
    
    \textbf{Answer to RQ\textsubscript{2}}:
    \emph{Class containers (context vector), lexical metrics, and systematically constructed metrics improve the prediction performance of the machine learning models measuring testability. These metrics improve the model mean absolute error by an average of 0.0571 (47.94\%).
}
\end{tcolorbox}

\subsection{Influential testability metrics}\label{sec:metrics-analysis}
\begin{figure}
    \centering
    \includegraphics[width=1\linewidth]{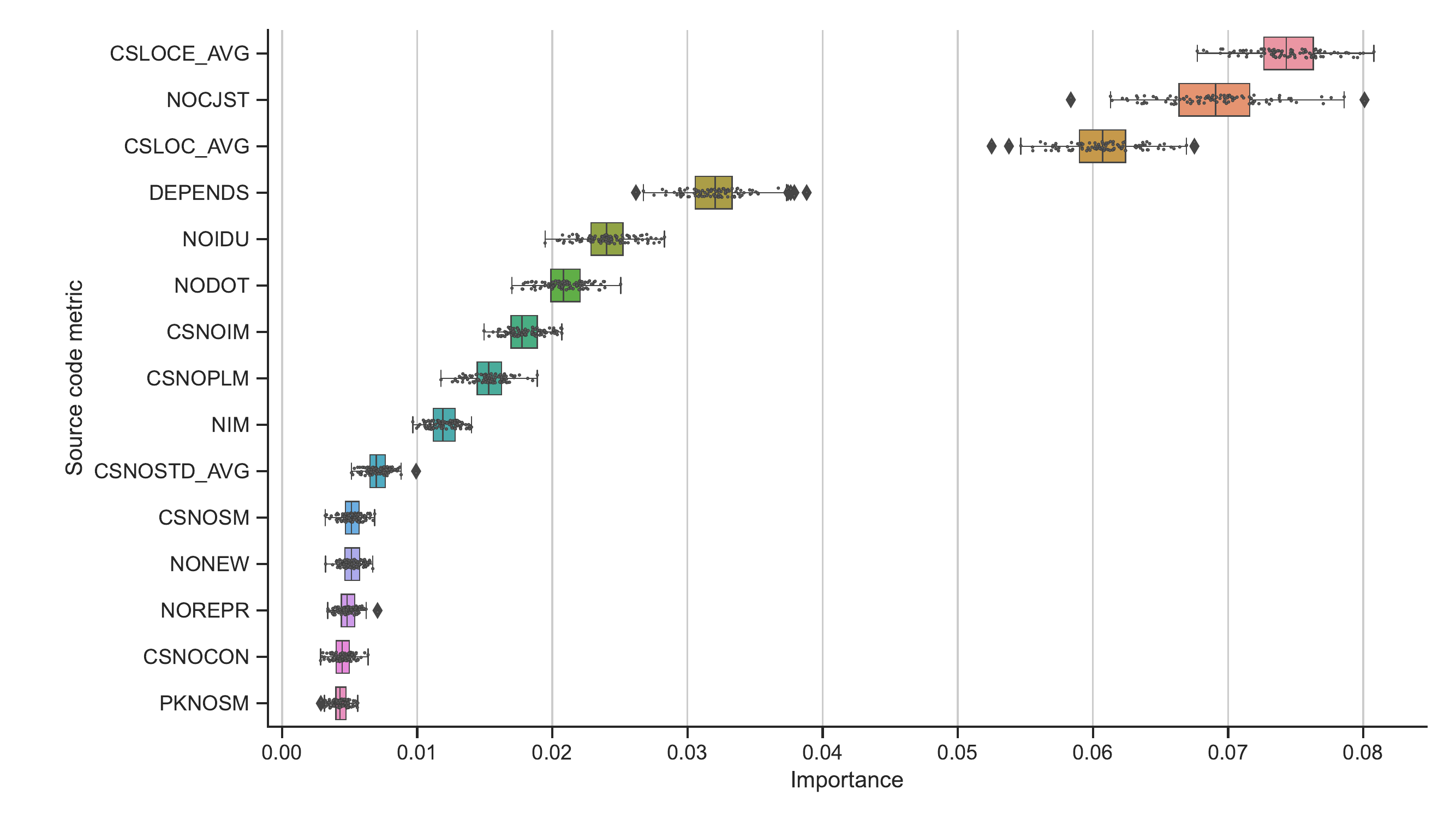}
    \caption{Top 15 influential source code metrics in predicting testability.}
    \label{fig:permutation-importance}
\end{figure}

\noindent
To answer RQ\textsubscript{3}, the importance of each metric for the best model, VoR, was computed on the best dataset, DS1. The permutation importance technique \cite{Breiman2001} was used since the VoR model, unlike tree-based models, lacks any built-in feature importance functionality. In this technique, the values of a single feature are shuffled, and the learned model is asked to make predictions using the same test set with the shuffled feature. 
Comparing these predictions and the actual target values determines the extent to which the learned model is affected by shuffling. The performance deterioration denotes the importance of the shuffled feature, which is a source code metric in the testability prediction test set.
The permutation importance process was repeated 100 times to alleviate the effects caused by the stochastic nature of this technique.
Figure \ref{fig:permutation-importance} shows the box-plot of changes to the model's R\textsuperscript{2} score after repeated permutation for the top 15 influential source code metrics. The following results are observed:

\begin{itemize}
\item{The average lines of execution codes of the class under test (CSLOCE\_AVG) is the most influential metric affecting class testability. Removing this metric reduces the R\textsuperscript{2} score of the prediction model by 8\%.}

\item{Important metrics are scattered among all quality subjects, including size, complexity, coupling, visibility, and inheritance. It means that source code testability is affected by all various software quality aspects.}

\item{\emph{Nine} of the 15 selected essential metrics in Figure \ref{fig:permutation-importance} are the newly proposed metrics in this paper, including lexical metrics, sub-metrics, and package-level metrics. It confirms that vectorization of the source code with newly defined metrics improves the prediction of testability with machine learning models. 
}
\end{itemize}

\begin{figure}[!h]
    \centering
    \includegraphics[width=1.005\linewidth, keepaspectratio]{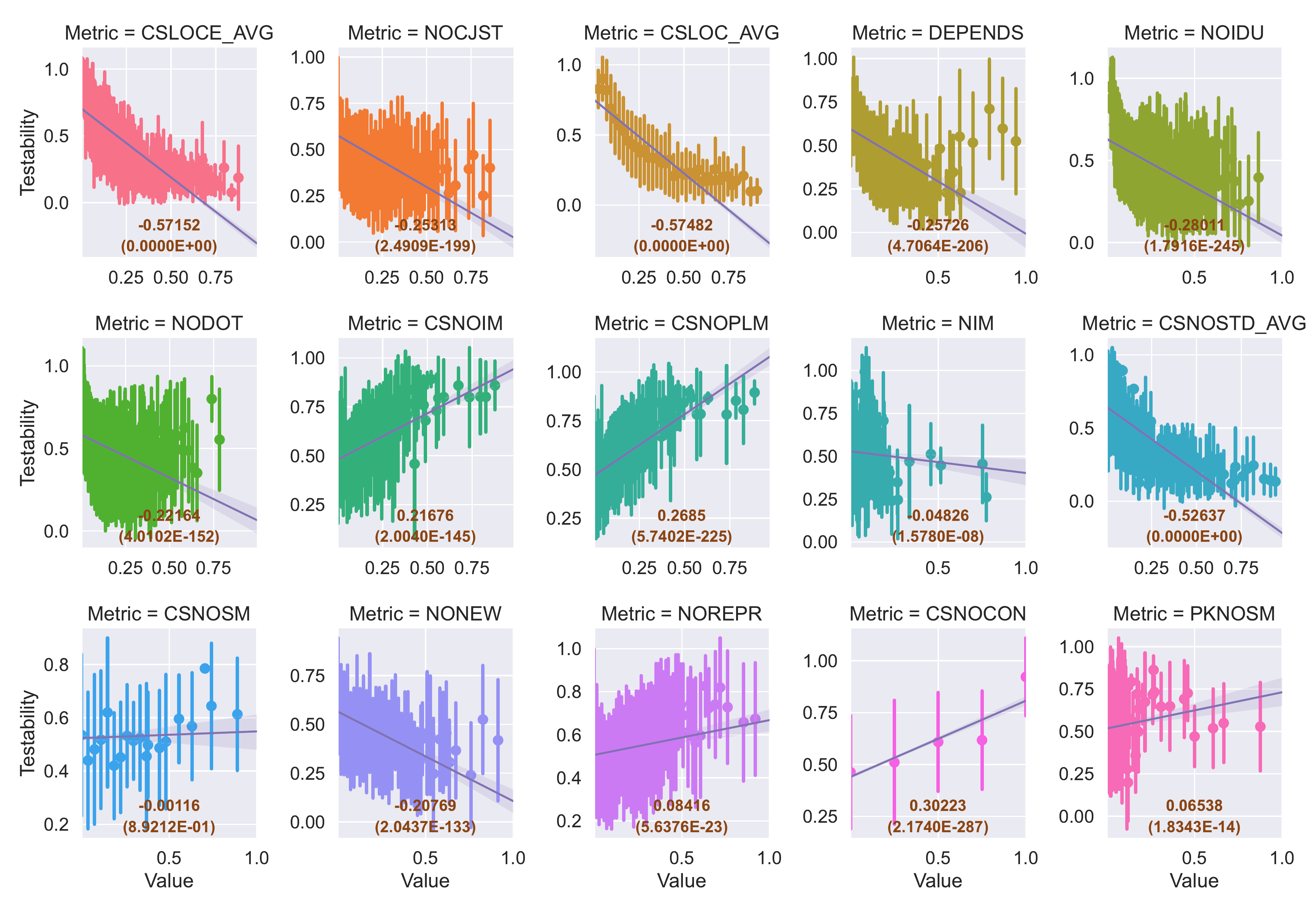}
    \caption{Correlation between testability and important predictors.}
    \label{fig:metrics-testability-relationship4.png}
\end{figure}

The type of correlation (positive or negative) between each influential metric and testability is essential to determine in which direction testability can be improved. The Pearson correlation analysis was used to find the relationship between testability and influential source code metrics. 
 Figure \ref{fig:metrics-testability-relationship4.png} shows the correlation between the testability value and the top 15 influential metrics. A normalized metric value is used to improve the visualization of source code metrics and regression lines.
 The Pearson correlation coefficient, along with the associated p-value, is shown on each plot in Figure \ref{fig:metrics-testability-relationship4.png}. A positive correlation indicates that increasing the value of an influential metric increases the testability, while a negative correlation means the opposite. The p-value indicates whether the calculated correlation is statistically significant or not. A p-value greater than 0.01 implies that the correlation is not statistically significant. 
  
Significant negative correlations are observed for nine out of 15 metrics, and positive correlations are observed for five out of 15 metrics. Moreover, the correlation for one metric, the number of static methods (CSNOSM), is not significant. Indeed, most source code metrics, as already expected, negatively impact testability. However, the class instance methods (CSNOIM), public methods (CSNOPLM), print and return statements (CSNOREPR), constructors (CSNOCON), and static methods in its enclosing package (PKNOSM) increase the class testability.

Investigation of the CSNOIM metric reveals that classes with many small methods are more testable than classes with few methods in which the methods are typically long. The analysis of other metrics positively correlates with testability, including CSNOPLM, CSNOREPR, and CSNOCON, indicating that these metrics primarily measure the visibility of the class. 
 For instance, the outputs of methods with return statements are more observable and testable than the method which implicitly modified many class fields. Related works have shown the positive impact of visibility and observability measures on testability \cite{ Voas1995, Ma2017}. Therefore, the findings in this research sightly support the results of previous research.
 
 Another interesting observation is that the static methods in the enclosing package of a class (PKNOSM) increase the class testability while static methods inside that class tend to decrease its testability. It concludes that defining static methods in a class is not desirable for testability. However, once they are defined, using them in other classes increases testability. The main reason is that calling static methods requires no object instantiation and constructor invocation, which reduces the coupling between objects.

Overall, classes with few lines of code, few jump statements, and many visible methods are more testable than other classes.
These results provide informative clues about the automatic modification of source code to enhance testability. 
Some empirical evidence on the possibility of testability improvement by automated refactoring is demonstrated in the next section.
 
\begin{tcolorbox}[boxrule=0.5pt]
    \textbf{RQ\textsubscript{3}}:
    \textit{Which source code metrics affect the testability of a class more than others?}
    
    \textbf{Answer to RQ\textsubscript{3}}: 
    \emph{Feature importance analysis automatically determines the critical source code metrics affecting testability without any human interventions. This set of metrics includes lines of code, conditional jump statements, class dependencies, and the number of identifiers defined in the class. 
}
\end{tcolorbox}

\subsection{Testability improvement}
\noindent
The impact of automated refactoring on source code testability has remained an open question \cite{Elish2009, Alshayeb2009, Cinneide2011}. Testability prediction provides a means to systematically investigate this question by determining the source code metrics that should be focused on when refactoring.
Table \ref{tabel:selected-refactorings} presents a set of refactorings, directly improving those source code metrics that affect testability.

\begin{table}[]
    \centering
    \caption{Refactoring operations that potentially improve testability.}
    \label{tabel:selected-refactorings}
    \resizebox{0.85\linewidth}{!}{%
        \begin{tabular}{ll} 
            \hline
            Source code metrics  & 
            Refactoring   
             \\  \hline
            \begin{tabular}[c]{@{}l@{}}
                LOC, NOST, NOIDU, NODOT, NONEW
            \end{tabular} &
             \begin{tabular}[c]{@{}l@{}}
                 Extract class, move method, remove dead code
             \end{tabular}  \\
         \rowcolor[rgb]{0.985,0.985,0.985} NOIM, NOCON &
         Extract method, move method
          \\
            NOCJS  & 
            Simplify conditional logic, extract method   \\
            \rowcolor[rgb]{0.985,0.985,0.985}  DEPENDS    & 
            Move method, extract class  \\
            NOPLM  &
             Increase method visibility  \\
            \rowcolor[rgb]{0.985,0.985,0.985} NOSM   & 
            Make method non-static  \\
            NOREPR & 
             Extract method, remove dead code   \\
            \rowcolor[rgb]{0.985,0.985,0.985} NIM  & 
            Extract superclass, pull-up method   \\
            \hline
        \end{tabular}
    }
\end{table}

The least testable classes of the two well-known Java projects, Weka \cite{Hall2009} and Scijava-common \cite{Scijava-common2021}, with at least one automated refactoring opportunity, are chosen to be used in this experiment. These projects contain more than 1,000 classes with many scientific and statistical operations, which make their testing difficult and time-consuming.
The testability of the classes was measured before and after applying each automated refactoring by existing refactoring tools.
Three existing refactoring tools, JDeodorant \cite{Tsantalis2018}, MultiRefactor \cite{Mohan2017}, and IntelliJ IDEA \cite{IntelliJ-IDEA2021}, were used to identify and apply refactorings listed in Table \ref{tabel:selected-refactorings}. Each tool supports some refactoring operations.
 The 'extract class', 'extract method', and 'move method' refactorings were performed by JDeodorant  \cite{Tsantalis2018}.
MultiRefactor \cite{IntelliJ-IDEA2021} was used to identify and apply the 'make method non-static', 'increase method visibility', 'pull-up method', and 'extract superclass' refactoring operations. The 'remove dead codes' and 'simplifying conditional logic' refactorings were applied by IntelliJ IDEA \cite{IntelliJ-IDEA2021}. 

Table \ref{table:refactoring-result} shows the number of refactorings, testability values, total test time, and total prediction time for selected classes in the Weka and Scijava-common projects. 
An average testability improvement of 0.1257 (a relative improvement of 87.35\%) in the Weka smelly classes and 0.2405 (a relative improvement of 86.44\%) in the Scijava-common smelly classes are observed.
In Table \ref{table:refactoring-result}, the column titled 'Test time' lists the time taken to run EvoSuite to test each class after refactoring and computing testability with Equation \ref{eq:6}. Prediction time includes the static analysis to compute source code metrics and the execution of Algorithm \ref{alg:inference-testability}.
It is observed that the test time of Weka classes has been improved by 2850.95 minutes (a relative improvement of 99.82\%). In the same way, the test time of Scijava-common classes has been improved by 317.84 minutes (a relative improvement of 99.94\%) which consider a significant enhancement. 

 \begin{table}
     \centering
     \caption{Impact of automated refactoring on testability.}
     \label{table:refactoring-result}
     \resizebox{0.85\linewidth}{!}{%
         \begin{tabular}{lllllll} 
             \hline
             \multirow{2}{*}{Project} & \multirow{2}{*}{\begin{tabular}[c]{@{}l@{}}Selected \\classes\end{tabular}} & \multirow{2}{*}{Refactorings} & \multicolumn{2}{l}{Testability}                     &  \multirow{2}{*}{\begin{tabular}[c]{@{}l@{}}Test time\\~(minutes)\end{tabular}} & \multirow{2}{*}{\begin{tabular}[c]{@{}l@{}}Prediction time\\~(minutes)\end{tabular}}  \\
             &                                                                             &                               & Before & {\cellcolor[rgb]{0.980,0.980,0.980}}After  &                                                                                &                                                                                       \\ 
             \hline
             Weka                     & 31                                                                          & 476                           & 0.1439 & {\cellcolor[rgb]{0.980,0.980,0.980}}0.2696 & 2856                                                                           & 5.0456                                                                                \\
             Scijava-common           & 11                                                                          & 53                            & 0.2782 & {\cellcolor[rgb]{0.980,0.980,0.980}}0.5187 & 318                                                                            & 0.1643                                                                                \\
             \hline
         \end{tabular}
     }
 \end{table}
 
\begin{figure}
    \centering
    \includegraphics[width=1\linewidth]{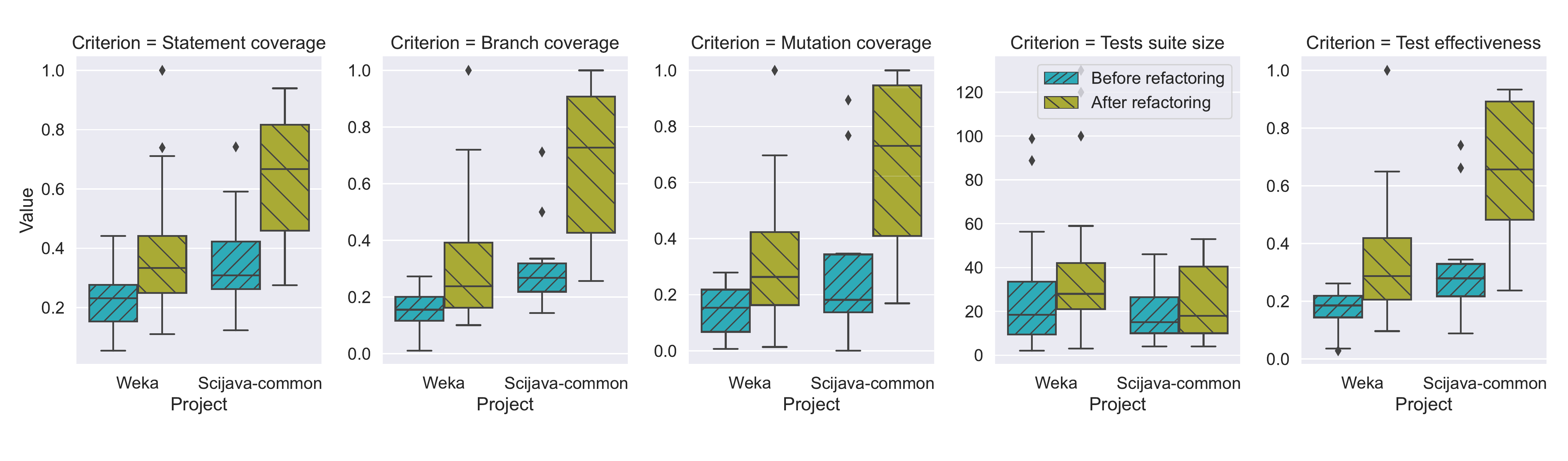}
    \caption{Distribution of important testability metrics.}
    \label{fig:test-effectiveness.png}
\end{figure}

 Figure \ref{fig:test-effectiveness.png} shows the coverage criteria, test suite size, and test effectiveness for the Weka and Scijava-common classes before and after refactoring. 
 Coverage criteria have been computed by running EvoSuite \cite{Arcuri2016} on each class five times and averaging the results. Test effectiveness has been computed by Equation \ref{eq:2}.
It is observed that all the coverage measures have been improved, and the size of the test suite also have increased while the test budget has been kept fixed.
Indeed, after refactoring, the number of influential tests grows, leading to increased test effectiveness and testability. 
 The test effectiveness of the Weka and Scijava-common classes has been improved by an average of 0.1585 (92.23\%) and 0.3228 (98.76\%), respectively. 
 
 The t-test statistical hypothesis test with a 99\% confidence level ($\alpha=0.01$) was employed to indicate whether the differences in the mean of different test criteria before and after refactoring are statistically significant or not. 
The null hypothesis (H0) was that the mean of the coverage provided by any test suite does not vary before and after refactoring. The alternative hypothesis (H1) was that the mean coverage increases after refactoring. 
 The p-value of the t-test corresponds to the probability of rejecting the null hypothesis (H0) while it is true (type I error). A p-value less than or equal to $\alpha\ (\le 0.01)$ means that the H1 is accepted and H0 is rejected. However, a p-value strictly greater than $\alpha\ (> 0.01)$ implies the opposite. 

Table \ref{table:code-coverage-t-test} shows the p-value of the t-test on refactored classes for each criterion. It is observed that the null hypothesis is rejected in all tests p-values strictly lower than $0.01$. It indicates that the test adequacy criteria and effectiveness increase significantly after applying relevant refactoring to classes with low testability.

\begin{table}[]
    \centering
    \caption{Results of t-test (p-values) on refactored classes for test criteria, before and after refactoring.}
    \label{table:code-coverage-t-test}
    \arrayrulecolor{black}
    \resizebox{0.90\linewidth}{!}{%
        \begin{tabular}{lllll} 
            \hline
            \multirow{2}{*}{Project} & \multicolumn{4}{l}{Independent t-test p-value}                                                          \\ 
            \arrayrulecolor{black}\cline{2-5}
            & Line coverage                               & Branch coverage & Mutation coverage & Test effectiveness  \\ 
            \hline
            Weka                     & \begin{tabular}[c]{@{}l@{}}$1.3640\times{10}^{-4}$\\\end{tabular} & $4.8578\times{10}^{-5}$               & $1.7451\times{10}^{-4}$                & $3.3348\times{10}^{-5}$                   \\
            Scijava-common           & $3.1345\times{10}^{-3}$                                          & $9.7284\times{10}^{-4}$              & $5.7722\times{10}^{-3}$                 & $1.9778\times{10}^{-3}$                   \\
           \rowcolor[rgb]{0.920,0.920,0.950} All                      & $3.4405\times{10}^{-5}$                                         & $1.4134\times{10}^{-5}$            & $6.6132\times{10}^{-5}$             & $1.3586\times{10}^{-5}$                \\
            \arrayrulecolor{black}\hline
        \end{tabular}
    }
\end{table}
 
\begin{figure}[!h]
    \centering
    \includegraphics[width=1\linewidth]{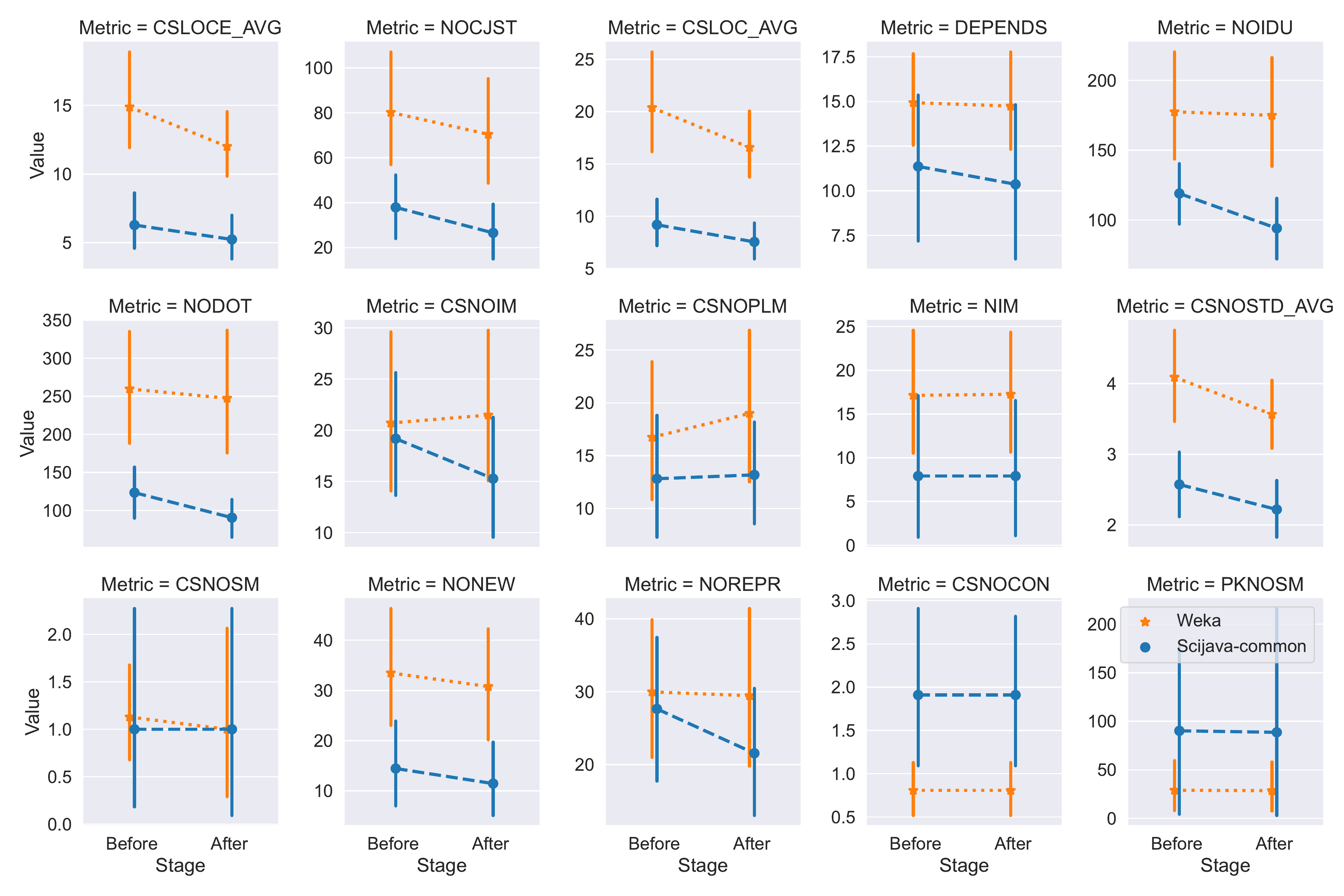}
    \caption{Changes in the source code metrics (before and after refactoring).}
    \label{fig:metrics-before-after-refactoring}
\end{figure}

Figure \ref{fig:metrics-before-after-refactoring} shows the impact of refactoring on the source code metrics affecting testability. 
By comparing Figure \ref{fig:metrics-testability-relationship4.png} and Figure \ref{fig:metrics-before-after-refactoring}, it is observed that the applied refactorings change source code metrics such that the testability prediction model could predict a higher testability value for the refactored class.

\begin{tcolorbox}[boxrule=0.5pt]
    \textbf{RQ\textsubscript{4}}:
    \textit{Is it possible to improve software testability by improving influential source code metrics via automatic refactoring?}
    
    \textbf{Answer to RQ\textsubscript{4}}:
    \emph{Refactorings focused on source code metrics that affect testability prediction improved the testability of 42 Java low testable classes by an average of 0.1831 (86.87\%). Testability improvement significantly enhances test adequacy criteria, including statement, branch, and mutation coverage of the class under test. It should be noted that predicting testability rather than actually testing the code before and after refactoring saved the time caused due to extra testing by about 99.89\%.
    }.
\end{tcolorbox}

\subsection{Evaluating other quality attributes}
Testability is only one property of the code. Being able to read, understand, and maintain the code easily are also very important. The impact of refactoring for testability enhancement on the other quality factors is investigated to answer RQ\textsubscript{5}.
To this aim, in addition to testability, four other quality attributes, including reusability, functionality, extensibility, and modularity for the Weka \cite{Hall2009} and Scijava-common \cite{Scijava-common2021} classes, were measured before refactoring for testability enhancement. The first three attributes are measured using the relations offered in \cite{Bansiya2002}. Modularity was computed using an approach based on the concept of modularity in complex networks \cite{Leicht2008, Xiang2019}.
Table \ref {table:quality-attrs} shows the definitions and computation equations of these quality attributes.
The reusability, functionality, and extensibility attributes are computed using QMOOD design quality metrics \cite{Bansiya2002}.
Modularity, $Q$, is computed considering the module dependency graph (MDG) \cite{Mitchell2006}. MDG includes all of the modules in the system and the set of dependencies that exist between the modules.
In the modularity equation, $A$ is the adjacency matrix of the MDG; $m$ is the number of components (e.g., packages in Java programs); $k_i^{in}$ is the input degree of node $i$ (e.g., a class in Java programs) in the MDG; $k_j^{out}$ is the output degree of node $j$ in the MDG; $c_{i}$ and $c_j$ are the modules that nodes $i$ and $j$ belong to; and $\delta$ is the Kronecker delta function that takes 1 when $c_i$ equals $c_j$, and 0, otherwise.

\begin{table}
    \centering
    \caption{Definitions and computation equations of quality attributes.}
    \label{table:quality-attrs}
    \arrayrulecolor{black}
    \resizebox{\linewidth}{!}{%
        \begin{tabular}{lll} 
            \hline
            Quality attribute & Definition                                                                                                                                                                                                  & Computation equation                                                                                                                                                                                                                                      \\ 
            \hline
            Reusability       & \begin{tabular}[c]{@{}l@{}}Reflects the presence of object-oriented design \\characteristics that allow a design to be reapplied  \\to a new problem without significant effort.\end{tabular}              & 
            \begin{tabular}[c]{@{}l@{}}
                $-0.25$ $\times$ class coupling + 0.25 $\times$ cohesion among methods \\in class + 0.5 $\times$ number of public methods in a class + \\ 0.5 $\times$ design size in classes
            \end{tabular}                                                                    \\
             \rowcolor[rgb]{0.985,0.985,0.985} Functionality     &
              \begin{tabular}[c]{@{}l@{}}
                  The responsibilities assigned to the classes of a  \\design, which are made available by the classes  \\through their public interfaces.\end{tabular}&
               \begin{tabular}[c]{@{}l@{}}0.12 $\times$ cohesion among methods in class + 0.22 $\times$ number  \\of polymorphic methods in a class + 0.22 $\times$ number of  \\public methods in a class + 0.22 $\times$ design size in classes  \\+ 0.22 $\times$ number of hierarchies
              \end{tabular}  \\
            Extendibility     & 
            \begin{tabular}[c]{@{}l@{}}
                Refers to the presence and usage of properties in \\ an existing design that incorporates new \\  requirements in the design.
            \end{tabular}    &
         \begin{tabular}[c]{@{}l@{}}
             0.5 $\times$  average number of ancestors $ - 0.5$ $\times$ class coupling  \\+ 0.5 $\times$ number of inherited methods in a class + 0.5  $\times$ \\number of polymorphic methods in a class
         \end{tabular}                                                    \\
             \rowcolor[rgb]{0.985,0.985,0.985} Modularity        & \begin{tabular}[c]{@{}l@{}}Degree to which a system or computer program is  \\composed of discrete components such that a  \\change to one component has minimal impact \\on other components.\end{tabular} & 
            $
             Q =  \frac{1}{m} \sum_{ij}^{} \left(A_{ij} - \frac{k_{i}^{in} \times k_{j}^{out}}{m}\right) \delta(c_i, c_j)
            $                                                                                                                                          \\
            \hline
        \end{tabular}
    }
    \arrayrulecolor{black}
\end{table}

\begin{figure*}
    \centering
    \includegraphics[width=1\linewidth]{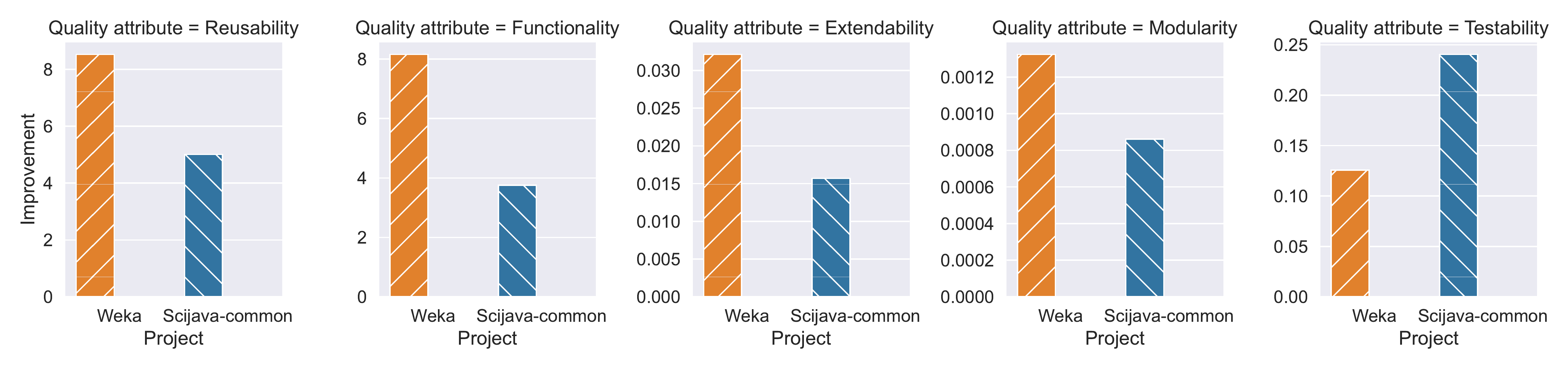}
    \caption{Improvement of different quality attributes along with testability.}
    \label{fig:qmood-modularity-testability}
\end{figure*}

Figure \ref{fig:qmood-modularity-testability} shows the improvement of the aforementioned quality attributes after refactoring classes of Weka and Scijava-common projects. It is observed that all quality attributes are improved after the refactorings. 
Figure \ref{fig:qmood-modularity-testability} indicates an average of 6.76, 5.96, 0.02, and 0.001 improvements in the reusability, functionality, extendability, and modularity quality attributes of the Weka and SciJava-Common software systems.

It is worth mentioning that testability improvement, together with other quality attributes, is not already mentioned by the other researchers \cite{Mkaouer2016, Mohan2019}.
All the experiments were performed with automatic test data generation methods. At this point, improving other quality attributes besides testability assures that even manual testing could also be more straightforward after refactoring.

\begin{tcolorbox}[boxrule=0.5pt]
    \textbf{RQ\textsubscript{5}}:
    \textit{Does refactoring for testability improve other quality attributes?}

     \textbf{Answer to RQ\textsubscript{5}}:
      \emph{Automated refactoring for testability improves other quality attributes, including reusability, functionality, extendability, and modularity, which is useful for manual testing.}
\end{tcolorbox}

\section{Threats to validity}\label{sec:threats-to-validity}
\noindent 
Threats to construct validity are on how the testability model is defined. In this paper, the standard testability definition was used to build a new software testability measurement model, which emphasizes test quality and effort metrics. Test adequacy criteria,\textit{i.e.}, statement, branch, and mutation coverage, were computed with a fixed test budget for all the projects to accomplish a fair measurement. However, other test quality and effort metrics can contribute to the testability measurement model, Equation \ref{eq:6},  to achieve more realistic results. 
Test adequacy criteria, and on top of that, coverage criteria are general concepts not restricted to any programming language. Therefore, the proposed mathematical model can be used for any programming language. However, the source code metrics used to vectorize or represent source code components may vary depending on whether the programming language is object-oriented, domain-specific, functional, or logical. The metrics introduced in this paper, including lexical metrics and sub-metrics, are independent of the programming languages' type and syntax.  

The most important threats to internal validity are the random nature of both evolutionary and machine learning algorithms used to compute and learn testability. For the sake of the reliability of the coverage provided by EvoSuite tests, the test suite generation process was repeated five times for each project. The EvoSuite tool with default settings and different random seeds was used to generate five test suites. Thereafter, the results obtained for each criterion were averaged. 
A low variance of results was observed on average for all projects, which minimizes the possibility of randomness in the results. However, there is still an opportunity to test with different hyperparameters. 
The grid search strategy, along with the five-fold cross-validation, was used to ensure the reliability of the training process, prevent overfitting, and produce the best possible machine learning model for testability prediction. 
In addition, the training of each model was performed with different random seeds, and the prediction results were averaged to minimize the randomness of the results.

The main threat to external validity regards the testability prediction models' application to the programs in other programming languages rather than Java. 
The SF110 corpus \cite{Fraser2014} containing 110 Java projects was used to ensure a generalization of all software types. The proposed model can be used to predict the testability of programs written in other programming languages. Nevertheless, there is a need to evaluate this approach on different programming languages to ensure the goodness of metrics and models.

\section{Conclusion}\label{sec:conclusion}
\noindent 
Testability is effectively proportional to the number of test data required to achieve a certain degree of code coverage. The fewer the test data needed to achieve the desired code coverage, the more testable the code. Experimental results in this paper demonstrate an average of 95.5\% improvement in coverage achieved by the test data generated in 252 minutes after improving the testability of 42 Java classes by an average of 86.87\%. In addition to improving test effectiveness, testability prediction helps testers avoid ineffective and unnecessary tests. By predicting the amount of testability instead of generating and measuring the test suite coverage after each refactoring, the time required for test-based development and modification can be significantly reduced. The empirical studies in this article demonstrate a 99.89\% improvement in the time required to improve the testability of Weka and Scijava-common software systems from 0.12 and 0.27 to 0.27 and 0.52, respectively. 

Testability prediction could be applied at any stage of software development, even before the code under test is ready to execute. The proposed model in this paper achieves an R\textsuperscript{2}-score of 0.68, which improves the performance of existing testability prediction models by 43\%. The plausible improvement in the performance is due to the increased number of source code metrics and the data samples used to learn the proposed model. 

This paper designates 15 software metrics highly affecting testability. The metrics can be identified automatically by applying features importance analysis. Improvements of these metrics, by automated refactoring, may significantly improve test adequacy criteria, specifically statement coverage, branch coverage, and mutation score for the units under test. It is shown in this paper that testability improvement also improves other software quality attributes, including reusability, functionality, extendability, and modularity. Experimental results in this paper indicate an average of 6.76, 5.96, 0.02, and 0.001 improvements in the reusability, functionality, extendability, and modularity quality attributes of the Weka and Scijava-common software systems. 

In future work, the authors aim to introduce a new software development methodology, testability-driven development (TsDD), to substitute test-driven development (TDD) in agile methodologies. Testing by itself could be a lengthy and costly process, specifically if the code under test solves a scientific formula or controls a cyber-physical system. TDD aggravates the cost by encouraging testing as a tool for the incremental development of software. TsDD suggests postponing the test to when the testability is optimized due to frequent refactoring and measuring. Testability prediction could also be used as an objective function in search-based processes to look for the sequence of refactorings maximizing testability.
 
\section*{Compliance with Ethical Standards}
\noindent This study has received no funding from any organization. 

\section*{Conflict of Interest}
\noindent The authors declare that they have no conflict of interest.

\section*{Ethical Approval}
\noindent This article does not contain any studies with human participants or animals performed by any of the authors.

\bibliographystyle{elsarticle-num}
\bibliography{refs}

\begin{thebibliography}{10}
\expandafter\ifx\csname url\endcsname\relax
  \def\url#1{\texttt{#1}}\fi
\expandafter\ifx\csname urlprefix\endcsname\relax\def\urlprefix{URL }\fi
\expandafter\ifx\csname href\endcsname\relax
  \def\href#1#2{#2} \def\path#1{#1}\fi

\bibitem{Dijkstra1972}
E.~W. Dijkstra, \href{https://doi.org/10.1145/355604.361591}{The humble
  programmer}, Commun. ACM 15~(10) (1972) 859–866.
\newblock \href {https://doi.org/10.1145/355604.361591}
  {\path{doi:10.1145/355604.361591}}.
\newline\urlprefix\url{https://doi.org/10.1145/355604.361591}

\bibitem{Ammann2016}
P.~Ammann, J.~Offutt, {Introduction to Software Testing}, Cambridge University
  Press, Cambridge, 2016.
\newblock \href {https://doi.org/DOI: 10.1017/9781316771273} {\path{doi:DOI:
  10.1017/9781316771273}}.

\bibitem{Khan2009}
R.~A. Khan, K.~Mustafa,
  \href{http://portal.acm.org/citation.cfm?doid=1507195.1507204}{{Metric based
  testability model for object oriented design (MTMOOD)}}, ACM SIGSOFT Software
  Engineering Notes 34~(2) (2009) 1.
\newblock \href {https://doi.org/10.1145/1507195.1507204}
  {\path{doi:10.1145/1507195.1507204}}.
\newline\urlprefix\url{http://portal.acm.org/citation.cfm?doid=1507195.1507204}

\bibitem{MuhammadRabeeShaheen2014}
L.~D.~B. {Muhammad Rabee Shaheen},
  \href{https://hal.inria.fr/hal-00953403}{{Survey of source code metrics for
  evaluating testability of object oriented systems}}, Tech. rep., Inria France
  (2014).
\newline\urlprefix\url{https://hal.inria.fr/hal-00953403}

\bibitem{Suri2015}
P.~R. Suri, H.~Singhani, {Object-oriented software testability (OOSTe) metrics
  analysis}, International Journal of Computer Applications Technology and
  Research 4~(5) (2015) 359--367.
\newblock \href {https://doi.org/10.7753/IJCATR0405.1006}
  {\path{doi:10.7753/IJCATR0405.1006}}.

\bibitem{Cohen1989}
I.~Cohen, T.~J. Mccabe, C.~W. Butler, {Design complexity measurement and
  testing} 32~(12) (1989) 1415--1423.

\bibitem{Chidamber1994}
S.~Chidamber, C.~Kemerer, \href{http://ieeexplore.ieee.org/document/295895/}{{A
  metrics suite for object oriented design}}, IEEE Transactions on Software
  Engineering 20~(6) (1994) 476--493.
\newblock \href {https://doi.org/10.1109/32.295895}
  {\path{doi:10.1109/32.295895}}.
\newline\urlprefix\url{http://ieeexplore.ieee.org/document/295895/}

\bibitem{Bieman1995}
J.~M. Bieman, B.-K. Kang,
  \href{http://portal.acm.org/citation.cfm?doid=223427.211856}{{Cohesion and
  reuse in an object-oriented system}}, ACM SIGSOFT Software Engineering Notes
  20~(SI) (1995) 259--262.
\newblock \href {https://doi.org/10.1145/223427.211856}
  {\path{doi:10.1145/223427.211856}}.
\newline\urlprefix\url{http://portal.acm.org/citation.cfm?doid=223427.211856}

\bibitem{Garousi2019}
V.~Garousi, M.~Felderer, F.~N. Kılı{\c{c}}aslan,
  \href{https://linkinghub.elsevier.com/retrieve/pii/S0950584918302490}{{A
  survey on software testability}}, Information and Software Technology 108
  (2019) 35--64.
\newblock \href {https://doi.org/10.1016/j.infsof.2018.12.003}
  {\path{doi:10.1016/j.infsof.2018.12.003}}.
\newline\urlprefix\url{https://linkinghub.elsevier.com/retrieve/pii/S0950584918302490}

\bibitem{Sharma2018}
R.~Sharma, A.~Saha, \href{https://ieeexplore.ieee.org/document/8675006/}{{A
  systematic review of software testability measurement techniques}}, in: 2018
  International Conference on Computing, Power and Communication Technologies
  (GUCON), IEEE, 2018, pp. 299--303.
\newblock \href {https://doi.org/10.1109/GUCON.2018.8675006}
  {\path{doi:10.1109/GUCON.2018.8675006}}.
\newline\urlprefix\url{https://ieeexplore.ieee.org/document/8675006/}

\bibitem{IEEEStd610.12-1990}
Ieee standard glossary of software engineering terminology, IEEE Std
  610.12-1990 (1990) 1--84\href {https://doi.org/10.1109/IEEESTD.1990.101064}
  {\path{doi:10.1109/IEEESTD.1990.101064}}.

\bibitem{ISOandIEC2011}
{ISO and IEC}, \href{https://www.iso.org/standard/35733.html}{{ISO/IEC
  25010:2011 systems and software engineering — systems and software quality
  requirements and evaluation (SQuaRE) — system and software quality models}}
  (2011) 34.
\newline\urlprefix\url{https://www.iso.org/standard/35733.html}

\bibitem{Terragni2020}
V.~Terragni, P.~Salza, M.~Pezz{\`{e}},
  \href{https://dl.acm.org/doi/10.1145/3387904.3389273}{{Measuring software
  testability modulo test quality}}, in: Proceedings of the 28th International
  Conference on Program Comprehension, ACM, New York, NY, USA, 2020, pp.
  241--251.
\newblock \href {https://doi.org/10.1145/3387904.3389273}
  {\path{doi:10.1145/3387904.3389273}}.
\newline\urlprefix\url{https://dl.acm.org/doi/10.1145/3387904.3389273}

\bibitem{Bruntink2004}
M.~{Bruntink}, A.~{van Deursen}, Predicting class testability using
  object-oriented metrics, in: Source Code Analysis and Manipulation, Fourth
  IEEE International Workshop on, 2004, pp. 136--145.
\newblock \href {https://doi.org/10.1109/SCAM.2004.16}
  {\path{doi:10.1109/SCAM.2004.16}}.

\bibitem{Bruntink2006}
M.~Bruntink, A.~van Deursen,
  \href{https://doi.org/10.1016/j.jss.2006.02.036}{An empirical study into
  class testability}, J. Syst. Softw. 79~(9) (2006) 1219–1232.
\newblock \href {https://doi.org/10.1016/j.jss.2006.02.036}
  {\path{doi:10.1016/j.jss.2006.02.036}}.
\newline\urlprefix\url{https://doi.org/10.1016/j.jss.2006.02.036}

\bibitem{BadriLinda2011}
L.~Badri, M.~Badri, F.~Toure, An empirical analysis of lack of cohesion metrics
  for predicting testability of classes, International Journal of Software
  Engineering and Its Applications 5~(2) (2011) 69--85.

\bibitem{BadriMourad2012}
M.~Badri, F.~Toure,
  \href{http://www.scirp.org/journal/doi.aspx?DOI=10.4236/jsea.2012.57060}{Empirical
  analysis of object-oriented design metrics for predicting unit testing effort
  of classes}, Journal of Software Engineering and Applications 05~(07) (2012)
  513--526.
\newblock \href {https://doi.org/10.4236/jsea.2012.57060}
  {\path{doi:10.4236/jsea.2012.57060}}.
\newline\urlprefix\url{http://www.scirp.org/journal/doi.aspx?DOI=10.4236/jsea.2012.57060}

\bibitem{Oluwatosin2020}
O.-J. Oluwatosin, A.~Balogun, S.~Basri, A.~Akintola, A.~Bajeh, Object-oriented
  measures as testability indicators: an empirical study, Journal of
  Engineering Science and Technology 15 (2020) 1092--1108.

\bibitem{Panichella2020}
A.~Panichella, J.~Campos, G.~Fraser,
  \href{https://dl.acm.org/doi/10.1145/3387940.3392266}{Evosuite at the sbst
  2020 tool competition}, ACM, 2020, pp. 549--552.
\newblock \href {https://doi.org/10.1145/3387940.3392266}
  {\path{doi:10.1145/3387940.3392266}}.
\newline\urlprefix\url{https://dl.acm.org/doi/10.1145/3387940.3392266}

\bibitem{Zakeri2021}
M.~Zakeri-Nasrabadi, S.~Parsa,
  \href{https://onlinelibrary.wiley.com/doi/abs/10.1002/int.22722}{Learning to
  predict test effectiveness}, International Journal of Intelligent Systems
  n/a~(n/a).
\newblock \href
  {http://arxiv.org/abs/https://onlinelibrary.wiley.com/doi/pdf/10.1002/int.22722}
  {\path{arXiv:https://onlinelibrary.wiley.com/doi/pdf/10.1002/int.22722}},
  \href {https://doi.org/https://doi.org/10.1002/int.22722}
  {\path{doi:https://doi.org/10.1002/int.22722}}.
\newline\urlprefix\url{https://onlinelibrary.wiley.com/doi/abs/10.1002/int.22722}

\bibitem{Breiman2001}
L.~Breiman, \href{https://doi.org/10.1023/A:1010933404324}{{Random forests}},
  Machine Learning 45~(1) (2001) 5--32.
\newblock \href {https://doi.org/10.1023/A:1010933404324}
  {\path{doi:10.1023/A:1010933404324}}.
\newline\urlprefix\url{https://doi.org/10.1023/A:1010933404324}

\bibitem{Carvalho2019}
D.~V. Carvalho, E.~M. Pereira, J.~S. Cardoso,
  \href{https://www.mdpi.com/2079-9292/8/8/832}{Machine learning
  interpretability: a survey on methods and metrics}, Electronics 8~(8) (2019).
\newblock \href {https://doi.org/10.3390/electronics8080832}
  {\path{doi:10.3390/electronics8080832}}.
\newline\urlprefix\url{https://www.mdpi.com/2079-9292/8/8/832}

\bibitem{IEEE1990}
IEEE, Ieee standard glossary of software engineering terminology, IEEE Std
  610.12-1990 (1990) 1--84\href {https://doi.org/10.1109/IEEESTD.1990.101064}
  {\path{doi:10.1109/IEEESTD.1990.101064}}.

\bibitem{ISO2001}
ISO/IEC, \href{https://www.iso.org/standard/22749.html}{Iso/iec 9126-1:2001
  software engineering — product quality — part 1: Quality model} (2001).
\newline\urlprefix\url{https://www.iso.org/standard/22749.html}

\bibitem{Toure2018}
F.~Toure, M.~Badri, L.~Lamontagne,
  \href{http://link.springer.com/10.1007/s11334-017-0306-1}{{Predicting
  different levels of the unit testing effort of classes using source code
  metrics: a multiple case study on open-source software}}, Innovations in
  Systems and Software Engineering 14~(1) (2018) 15--46.
\newblock \href {https://doi.org/10.1007/s11334-017-0306-1}
  {\path{doi:10.1007/s11334-017-0306-1}}.
\newline\urlprefix\url{http://link.springer.com/10.1007/s11334-017-0306-1}

\bibitem{BadriMourad2019}
M.~Badri, L.~Badri, O.~Hachemane, A.~Ouellet,
  \href{http://link.springer.com/10.1007/s11334-019-00334-6}{{Measuring the
  effect of clone refactoring on the size of unit test cases in object-oriented
  software: an empirical study}}, Innovations in Systems and Software
  Engineering 15~(2) (2019) 117--137.
\newblock \href {https://doi.org/10.1007/s11334-019-00334-6}
  {\path{doi:10.1007/s11334-019-00334-6}}.
\newline\urlprefix\url{http://link.springer.com/10.1007/s11334-019-00334-6}

\bibitem{Voas1995}
J.~Voas, K.~Miller, \href{http://ieeexplore.ieee.org/document/382180/}{Software
  testability: the new verification}, IEEE Software 12 (1995) 17--28.
\newblock \href {https://doi.org/10.1109/52.382180}
  {\path{doi:10.1109/52.382180}}.
\newline\urlprefix\url{http://ieeexplore.ieee.org/document/382180/}

\bibitem{Binder1994}
R.~V. Binder, \href{http://dl.acm.org/citation.cfm?doid=182987.184077}{Design
  for testability in object-oriented systems}, Communications of the ACM 37
  (1994) 87--101.
\newblock \href {https://doi.org/10.1145/182987.184077}
  {\path{doi:10.1145/182987.184077}}.
\newline\urlprefix\url{http://dl.acm.org/citation.cfm?doid=182987.184077}

\bibitem{Goel2008}
A.~Goel, S.~C. Gupta, S.~K. Wasan, {COTT – a testability framework for
  object-oriented software testing}, World Academy of Science, Engineering and
  Technology, International Journal of Computer, Electrical, Automation,
  Control and Information Engineering 2 (2008) 4224--4231.

\bibitem{Gonzalez-Sanchez2010}
A.~Gonzalez-Sanchez, E.~Piel, H.-G. Gross, A.~J.~C. van Gemund,
  \href{http://ieeexplore.ieee.org/document/5676288/}{{Minimising the
  preparation cost of runtime testing based on testability metrics}}, in: 2010
  IEEE 34th Annual Computer Software and Applications Conference, IEEE, 2010,
  pp. 419--424.
\newblock \href {https://doi.org/10.1109/COMPSAC.2010.49}
  {\path{doi:10.1109/COMPSAC.2010.49}}.
\newline\urlprefix\url{http://ieeexplore.ieee.org/document/5676288/}

\bibitem{Salahirad2019}
A.~Salahirad, H.~Almulla, G.~Gay,
  \href{https://onlinelibrary.wiley.com/doi/abs/10.1002/stvr.1701}{{Choosing
  the fitness function for the job: Automated generation of test suites that
  detect real faults}}, Software Testing, Verification and Reliability 29~(4-5)
  (jun 2019).
\newblock \href {https://doi.org/10.1002/stvr.1701}
  {\path{doi:10.1002/stvr.1701}}.
\newline\urlprefix\url{https://onlinelibrary.wiley.com/doi/abs/10.1002/stvr.1701}

\bibitem{Ma2017}
L.~Ma, C.~Zhang, B.~Yu, H.~Sato,
  \href{http://link.springer.com/10.1007/s11219-016-9340-8}{{An empirical study
  on the effects of code visibility on program testability}}, Software Quality
  Journal 25~(3) (2017) 951--978.
\newblock \href {https://doi.org/10.1007/s11219-016-9340-8}
  {\path{doi:10.1007/s11219-016-9340-8}}.
\newline\urlprefix\url{http://link.springer.com/10.1007/s11219-016-9340-8}

\bibitem{Daniel2008}
B.~Daniel, M.~Boshernitsan,
  \href{http://ieeexplore.ieee.org/document/4639342/}{{Predicting effectiveness
  of automatic testing tools}}, in: 2008 23rd IEEE/ACM International Conference
  on Automated Software Engineering, IEEE, 2008, pp. 363--366.
\newblock \href {https://doi.org/10.1109/ASE.2008.49}
  {\path{doi:10.1109/ASE.2008.49}}.
\newline\urlprefix\url{http://ieeexplore.ieee.org/document/4639342/}

\bibitem{Ferrer2013}
J.~Ferrer, F.~Chicano, E.~Alba,
  \href{https://www.sciencedirect.com/science/article/pii/S0950584913001535}{Estimating
  software testing complexity}, Information and Software Technology 55~(12)
  (2013) 2125--2139.
\newblock \href {https://doi.org/https://doi.org/10.1016/j.infsof.2013.07.007}
  {\path{doi:https://doi.org/10.1016/j.infsof.2013.07.007}}.
\newline\urlprefix\url{https://www.sciencedirect.com/science/article/pii/S0950584913001535}

\bibitem{Kobayashi2011}
H.~Kobayashi, B.~Mark, W.~Turin,
  \href{https://books.google.com/books?id=DQCMdT-3qbQC}{Probability, random
  processes, and statistical analysis: applications to communications, signal
  processing, queueing theory and mathematical finance}, Cambridge University
  Press, 2011.
\newline\urlprefix\url{https://books.google.com/books?id=DQCMdT-3qbQC}

\bibitem{Grano2019}
G.~Grano, T.~V. Titov, S.~Panichella, H.~C. Gall,
  \href{https://onlinelibrary.wiley.com/doi/abs/10.1002/smr.2158}{{Branch
  coverage prediction in automated testing}}, Journal of Software: Evolution
  and Process 31~(9) (sep 2019).
\newblock \href {https://doi.org/10.1002/smr.2158}
  {\path{doi:10.1002/smr.2158}}.
\newline\urlprefix\url{https://onlinelibrary.wiley.com/doi/abs/10.1002/smr.2158}

\bibitem{Hampel2011}
F.~R. Hampel, E.~M. Ronchetti, P.~J. Rousseeuw, W.~A. Stahel, {Robust
  statistics: the approach based on influence functions}, Wiley Series in
  Probability and Statistics, Wiley, 2011.

\bibitem{Chih-Chung2011}
C.-C. Chang, C.-J. Lin, \href{https://doi.org/10.1145/1961189.1961199}{Libsvm:
  A library for support vector machines}, ACM Trans. Intell. Syst. Technol.
  2~(3) (May 2011).
\newblock \href {https://doi.org/10.1145/1961189.1961199}
  {\path{doi:10.1145/1961189.1961199}}.
\newline\urlprefix\url{https://doi.org/10.1145/1961189.1961199}

\bibitem{Goodfellow2016}
I.~Goodfellow, Y.~Bengio, A.~Courville,
  \href{http://www.deeplearningbook.org/}{{Deep learning}}, MIT Press, 2016.
\newline\urlprefix\url{http://www.deeplearningbook.org/}

\bibitem{Noorian2011}
M.~Noorian, E.~B. Bagheri, W.~Du,
  \href{http://www.scopus.com/inward/record.url?eid=2-s2.0-84855543079&partnerID=40&md5=3224355016f222d0cb0b1a0e14111908}{Machine
  learning-based software testing: towards a classification framework}, SEKE
  2011 - Proceedings of the 23rd International Conference on Software
  Engineering and Knowledge Engineering (2011) 225--229.
\newline\urlprefix\url{http://www.scopus.com/inward/record.url?eid=2-s2.0-84855543079&partnerID=40&md5=3224355016f222d0cb0b1a0e14111908}

\bibitem{Zakeri2020}
M.~Z. Nasrabadi, S.~Parsa, A.~Kalaee, Format-aware learn\&fuzz: deep test data
  generation for efficient fuzzing, Neural Computing and Applications 33
  (2021).
\newblock \href {https://doi.org/10.1007/s00521-020-05039-7}
  {\path{doi:10.1007/s00521-020-05039-7}}.

\bibitem{Abdi2015}
Y.~Abdi, S.~Parsa, Y.~Seyfari,
  \href{https://doi.org/10.1007/s11334-015-0258-2}{A hybrid one-class rule
  learning approach based on swarm intelligence for software fault prediction},
  Innovations in Systems and Software Engineering 11 (2015) 289--301.
\newblock \href {https://doi.org/10.1007/s11334-015-0258-2}
  {\path{doi:10.1007/s11334-015-0258-2}}.
\newline\urlprefix\url{https://doi.org/10.1007/s11334-015-0258-2}

\bibitem{Shi2020}
K.~Shi, Y.~Lu, J.~Chang, Z.~Wei, Pathpair2vec: An ast path pair-based code
  representation method for defect prediction, Journal of Computer Languages 59
  (2020) 100979.
\newblock \href {https://doi.org/10.1016/J.COLA.2020.100979}
  {\path{doi:10.1016/J.COLA.2020.100979}}.

\bibitem{Mesquita2016}
D.~P. Mesquita, L.~S. Rocha, J.~P.~P. Gomes, A.~R.~R. Neto, Classification with
  reject option for software defect prediction, Applied Soft Computing 49
  (2016) 1085--1093.
\newblock \href {https://doi.org/10.1016/J.ASOC.2016.06.023}
  {\path{doi:10.1016/J.ASOC.2016.06.023}}.

\bibitem{Maru2019}
A.~Maru, A.~Dutta, K.~V. Kumar, D.~P. Mohapatra,
  \href{http://link.springer.com/10.1007/s12065-019-00318-2}{Software fault
  localization using bp neural network based on function and branch coverage},
  Evolutionary Intelligence (11 2019).
\newblock \href {https://doi.org/10.1007/s12065-019-00318-2}
  {\path{doi:10.1007/s12065-019-00318-2}}.
\newline\urlprefix\url{http://link.springer.com/10.1007/s12065-019-00318-2}

\bibitem{Dutta2021}
A.~Dutta, S.~S. Srivastava, S.~Godboley, D.~P. Mohapatra, Combi-fl: Neural
  network and sbfl based fault localization using mutation analysis, Journal of
  Computer Languages 66 (2021) 101064.
\newblock \href {https://doi.org/10.1016/J.COLA.2021.101064}
  {\path{doi:10.1016/J.COLA.2021.101064}}.

\bibitem{Alon20191}
U.~Alon, M.~Zilberstein, O.~Levy, E.~Yahav,
  \href{https://dl.acm.org/doi/10.1145/3290353}{code2vec: learning distributed
  representations of code}, Proceedings of the ACM on Programming Languages 3
  (2019) 1--29.
\newblock \href {https://doi.org/10.1145/3290353} {\path{doi:10.1145/3290353}}.
\newline\urlprefix\url{https://dl.acm.org/doi/10.1145/3290353}

\bibitem{Alon2018}
U.~Alon, S.~Brody, O.~Levy, E.~Yahav, Code2seq: generating sequences from
  structured representations of code, arXiv preprint arXiv:1808.01400 (2018).

\bibitem{Xiao2020}
H.~Xiao, M.~Cao, R.~Peng, Artificial neural network based software fault
  detection and correction prediction models considering testing effort,
  Applied Soft Computing 94 (2020) 106491.
\newblock \href {https://doi.org/10.1016/J.ASOC.2020.106491}
  {\path{doi:10.1016/J.ASOC.2020.106491}}.

\bibitem{Arcuri2016}
A.~Arcuri, J.~Campos, G.~Fraser, {Unit Test Generation During Software
  Development: EvoSuite Plugins for Maven, IntelliJ and Jenkins}, in: 2016 IEEE
  International Conference on Software Testing, Verification and Validation
  (ICST), 2016, pp. 401--408.
\newblock \href {https://doi.org/10.1109/ICST.2016.44}
  {\path{doi:10.1109/ICST.2016.44}}.

\bibitem{Fraser2014}
G.~Fraser, A.~Arcuri, \href{https://dl.acm.org/doi/10.1145/2685612}{{A
  large-scale evaluation of automated unit test generation using EvoSuite}},
  ACM Transactions on Software Engineering and Methodology 24~(2) (2014) 1--42.
\newblock \href {https://doi.org/10.1145/2685612} {\path{doi:10.1145/2685612}}.
\newline\urlprefix\url{https://dl.acm.org/doi/10.1145/2685612}

\bibitem{PachecoE2007Poster}
C.~Pacheco, M.~D. Ernst, {Randoop:} feedback-directed random testing for
  {Java}, in: OOPSLA 2007 Companion, Montreal, Canada, ACM, 2007.

\bibitem{NIPS2017}
G.~Ke, Q.~Meng, T.~Finley, T.~Wang, W.~Chen, W.~Ma, Q.~Ye, T.-Y. Liu,
  \href{http://papers.nips.cc/paper/6907-lightgbm-a-highly-efficient-gradient-boosting-decision-tree.pdf}{{LightGBM:
  A highly efficient gradient boosting decision tree}}, in: I.~Guyon, U.~V.
  Luxburg, S.~Bengio, H.~Wallach, R.~Fergus, S.~Vishwanathan, R.~Garnett
  (Eds.), Advances in Neural Information Processing Systems 30, Curran
  Associates, Inc., 2017, pp. 3146--3154.
\newline\urlprefix\url{http://papers.nips.cc/paper/6907-lightgbm-a-highly-efficient-gradient-boosting-decision-tree.pdf}

\bibitem{Guryanov2019}
A.~Guryanov, Histogram-based algorithm for building gradient boosting ensembles
  of piecewise linear decision trees, in: W.~M.~P. van~der Aalst, V.~Batagelj,
  D.~I. Ignatov, M.~Khachay, V.~Kuskova, A.~Kutuzov, S.~O. Kuznetsov, I.~A.
  Lomazova, N.~Loukachevitch, A.~Napoli, P.~M. Pardalos, M.~Pelillo, A.~V.
  Savchenko, E.~Tutubalina (Eds.), Analysis of Images, Social Networks and
  Texts, Springer International Publishing, Cham, 2019, pp. 39--50.

\bibitem{Fraser2013}
G.~Fraser, A.~Arcuri, Whole test suite generation, IEEE Transactions on
  Software Engineering 39~(2) (2013) 276--291.
\newblock \href {https://doi.org/10.1109/TSE.2012.14}
  {\path{doi:10.1109/TSE.2012.14}}.

\bibitem{Henderson-Sellers1995}
B.~Henderson-Sellers, {Object-oriented metrics: measures of complexity},
  Prentice-Hall, Inc., 1995.
\newblock \href {https://doi.org/Object-oriented metrics: measures of
  complexity} {\path{doi:Object-oriented metrics: measures of complexity}}.

\bibitem{Harrison1998}
R.~Harrison, S.~Counsell, R.~Nithi,
  \href{http://ieeexplore.ieee.org/document/689404/}{{An evaluation of the MOOD
  set of object-oriented software metrics}}, IEEE Transactions on Software
  Engineering 24~(6) (1998) 491--496.
\newblock \href {https://doi.org/10.1109/32.689404}
  {\path{doi:10.1109/32.689404}}.
\newline\urlprefix\url{http://ieeexplore.ieee.org/document/689404/}

\bibitem{Bansiya2002}
J.~Bansiya, C.~Davis, \href{http://ieeexplore.ieee.org/document/979986/}{{A
  hierarchical model for object-oriented design quality assessment}}, IEEE
  Transactions on Software Engineering 28~(1) (2002) 4--17.
\newblock \href {https://doi.org/10.1109/32.979986}
  {\path{doi:10.1109/32.979986}}.
\newline\urlprefix\url{http://ieeexplore.ieee.org/document/979986/}

\bibitem{ArcelliFontana2016}
F.~{Arcelli Fontana}, M.~V. M{\"{a}}ntyl{\"{a}}, M.~Zanoni, A.~Marino,
  \href{http://link.springer.com/10.1007/s10664-015-9378-4}{{Comparing and
  experimenting machine learning techniques for code smell detection}},
  Empirical Software Engineering 21~(3) (2016) 1143--1191.
\newblock \href {https://doi.org/10.1007/s10664-015-9378-4}
  {\path{doi:10.1007/s10664-015-9378-4}}.
\newline\urlprefix\url{http://link.springer.com/10.1007/s10664-015-9378-4}

\bibitem{NUNEZVARELA2017164}
A.~S. Nuñez-Varela, H.~G. Pérez-Gonzalez, F.~E. Martínez-Perez,
  C.~Soubervielle-Montalvo,
  \href{http://www.sciencedirect.com/science/article/pii/S0164121217300663}{Source
  code metrics: a systematic mapping study}, Journal of Systems and Software
  128 (2017) 164 -- 197.
\newblock \href {https://doi.org/https://doi.org/10.1016/j.jss.2017.03.044}
  {\path{doi:https://doi.org/10.1016/j.jss.2017.03.044}}.
\newline\urlprefix\url{http://www.sciencedirect.com/science/article/pii/S0164121217300663}

\bibitem{SciTools2020}
{SciTools}, {Understand}, [Online]. Available: \url{https://scitools.com/}
  ([Accessed: 2020-09-11]).

\bibitem{Breunig2000}
M.~M. Breunig, H.-P. Kriegel, R.~T. Ng, J.~Sander,
  \href{http://portal.acm.org/citation.cfm?doid=342009.335388}{{LOF:
  identifying density-based local outliers}}, in: Proceedings of the 2000 ACM
  SIGMOD international conference on Management of data - SIGMOD '00, ACM
  Press, New York, New York, USA, 2000, pp. 93--104.
\newblock \href {https://doi.org/10.1145/342009.335388}
  {\path{doi:10.1145/342009.335388}}.
\newline\urlprefix\url{http://portal.acm.org/citation.cfm?doid=342009.335388}

\bibitem{Scikit-learn2020}
F.~Pedregosa, G.~Varoquaux, A.~Gramfort, V.~Michel, B.~Thirion, O.~Grisel,
  M.~Blondel, P.~Prettenhofer, R.~Weiss, V.~Dubourg, J.~Vanderplas, A.~Passos,
  D.~Cournapeau, M.~Brucher, M.~Perrot, E.~Duchesnay, {Scikit-learn: machine
  learning in {\{}P{\}}ython}, Journal of Machine Learning Research 12 (2011)
  2825--2830.

\bibitem{Zhang2004}
T.~Zhang,
  \href{http://portal.acm.org/citation.cfm?doid=1015330.1015332}{{Solving large
  scale linear prediction problems using stochastic gradient descent
  algorithms}}, in: Twenty-first international conference on Machine learning -
  ICML '04, ACM Press, New York, New York, USA, 2004, p. 116.
\newblock \href {https://doi.org/10.1145/1015330.1015332}
  {\path{doi:10.1145/1015330.1015332}}.
\newline\urlprefix\url{http://portal.acm.org/citation.cfm?doid=1015330.1015332}

\bibitem{Bengio2012}
J.~Bergstra, Y.~Bengio, {Random search for hyper-parameter optimization}, J.
  Mach. Learn. Res. 13~(null) (2012) 281--305.

\bibitem{Cormen2022}
T.~H. Cormen, C.~E. Leiserson, R.~L. Rivest, C.~Stein,
  \href{https://mitpress.mit.edu/books/introduction-algorithms-fourth-edition}{Introduction
  to algorithms}, 4th Edition, MIT Press, 2022.
\newline\urlprefix\url{https://mitpress.mit.edu/books/introduction-algorithms-fourth-edition}

\bibitem{Aho2006}
A.~V. Aho, M.~S. Lam, R.~Sethi, J.~D. Ullman, Compilers: Principles,
  Techniques, and Tools (2nd Edition), Addison-Wesley Longman Publishing Co.,
  Inc., USA, 2006.

\bibitem{Xavier2014}
X.~Solé, A.~Ramisa, C.~Torras, Evaluation of random forests on large-scale
  classification problems using a bag-of-visual-words representation (2014).
\newblock \href {https://doi.org/10.3233/978-1-61499-452-7-273}
  {\path{doi:10.3233/978-1-61499-452-7-273}}.

\bibitem{Alpaydin2020}
E.~Alpaydin,
  \href{https://mitpress.mit.edu/books/introduction-machine-learning-fourth-edition}{Introduction
  to machine learning}, 4th Edition, MIT Press, 2020.
\newline\urlprefix\url{https://mitpress.mit.edu/books/introduction-machine-learning-fourth-edition}

\bibitem{Robinson2017}
R.~C. {da Cruz}., M.~{Medeiros Eler}., An empirical analysis of the correlation
  between ck metrics, test coverage and mutation score, in: Proceedings of the
  19th International Conference on Enterprise Information Systems - Volume 2:
  ICEIS,, INSTICC, SciTePress, 2017, pp. 341--350.
\newblock \href {https://doi.org/10.5220/0006312703410350}
  {\path{doi:10.5220/0006312703410350}}.

\bibitem{Elish2009}
K.~O. Elish, M.~Alshayeb,
  \href{http://ieeexplore.ieee.org/document/5358476/}{{Investigating the effect
  of refactoring on software testing effort}}, in: 2009 16th Asia-Pacific
  Software Engineering Conference, IEEE, 2009, pp. 29--34.
\newblock \href {https://doi.org/10.1109/APSEC.2009.14}
  {\path{doi:10.1109/APSEC.2009.14}}.
\newline\urlprefix\url{http://ieeexplore.ieee.org/document/5358476/}

\bibitem{Alshayeb2009}
M.~Alshayeb,
  \href{https://linkinghub.elsevier.com/retrieve/pii/S095058490900038X}{{Empirical
  investigation of refactoring effect on software quality}}, Information and
  Software Technology 51~(9) (2009) 1319--1326.
\newblock \href {https://doi.org/10.1016/j.infsof.2009.04.002}
  {\path{doi:10.1016/j.infsof.2009.04.002}}.
\newline\urlprefix\url{https://linkinghub.elsevier.com/retrieve/pii/S095058490900038X}

\bibitem{Cinneide2011}
M.~{\'{O}}. Cinn{\'{e}}ide, D.~Boyle, I.~H. Moghadam,
  \href{http://ieeexplore.ieee.org/document/5954444/}{{Automated Refactoring
  for Testability}}, in: 2011 IEEE Fourth International Conference on Software
  Testing, Verification and Validation Workshops, IEEE, 2011, pp. 437--443.
\newblock \href {https://doi.org/10.1109/ICSTW.2011.23}
  {\path{doi:10.1109/ICSTW.2011.23}}.
\newline\urlprefix\url{http://ieeexplore.ieee.org/document/5954444/}

\bibitem{Hall2009}
M.~Hall, E.~Frank, G.~Holmes, B.~Pfahringer, P.~Reutemann, I.~H. Witten,
  \href{http://doi.acm.org/10.1145/1656274.1656278
  http://portal.acm.org/citation.cfm?doid=1656274.1656278}{{The WEKA data
  mining software}}, ACM SIGKDD Explorations Newsletter 11~(1) (2009) 10.
\newblock \href {https://doi.org/10.1145/1656274.1656278}
  {\path{doi:10.1145/1656274.1656278}}.
\newline\urlprefix\url{http://doi.acm.org/10.1145/1656274.1656278
  http://portal.acm.org/citation.cfm?doid=1656274.1656278}

\bibitem{Scijava-common2021}
\href{https://github.com/scijava/scijava-common}{{Scijava-common}} ([Accessed:
  2021-06-23]).
\newline\urlprefix\url{https://github.com/scijava/scijava-common}

\bibitem{Tsantalis2018}
N.~Tsantalis, T.~Chaikalis, A.~Chatzigeorgiou,
  \href{http://ieeexplore.ieee.org/document/8330192/}{{Ten years of JDeodorant:
  Lessons learned from the hunt for smells}}, in: 2018 IEEE 25th International
  Conference on Software Analysis, Evolution and Reengineering (SANER), IEEE,
  2018, pp. 4--14.
\newblock \href {https://doi.org/10.1109/SANER.2018.8330192}
  {\path{doi:10.1109/SANER.2018.8330192}}.
\newline\urlprefix\url{http://ieeexplore.ieee.org/document/8330192/}

\bibitem{Mohan2017}
M.~Mohan, D.~Greer,
  \href{http://link.springer.com/10.1007/978-3-319-69926-4_46}{{MultiRefactor:
  automated refactoring to improve software quality}}, 2017, pp. 556--572.
\newblock \href {https://doi.org/10.1007/978-3-319-69926-4_46}
  {\path{doi:10.1007/978-3-319-69926-4_46}}.
\newline\urlprefix\url{http://link.springer.com/10.1007/978-3-319-69926-4_46}

\bibitem{IntelliJ-IDEA2021}
{JetBrains s.r.o.}, \href{https://www.jetbrains.com/idea/}{{IntelliJ IDEA}}
  ([Accessed: 2021-07-26]).
\newline\urlprefix\url{https://www.jetbrains.com/idea/}

\bibitem{Leicht2008}
E.~A. Leicht, M.~E.~J. Newman,
  \href{https://link.aps.org/doi/10.1103/PhysRevLett.100.118703}{Community
  structure in directed networks}, Physical Review Letters 100 (2008) 118703.
\newblock \href {https://doi.org/10.1103/PhysRevLett.100.118703}
  {\path{doi:10.1103/PhysRevLett.100.118703}}.
\newline\urlprefix\url{https://link.aps.org/doi/10.1103/PhysRevLett.100.118703}

\bibitem{Xiang2019}
Y.~Xiang, W.~Pan, H.~Jiang, Y.~Zhu, H.~Li,
  \href{https://www.mdpi.com/1099-4300/21/4/344}{{Measuring software modularity
  based on software networks}}, Entropy 21~(4) (2019) 344.
\newblock \href {https://doi.org/10.3390/e21040344}
  {\path{doi:10.3390/e21040344}}.
\newline\urlprefix\url{https://www.mdpi.com/1099-4300/21/4/344}

\bibitem{Mitchell2006}
B.~Mitchell, S.~Mancoridis,
  \href{http://ieeexplore.ieee.org/document/1610610/}{On the automatic
  modularization of software systems using the bunch tool}, IEEE Transactions
  on Software Engineering 32 (2006) 193--208.
\newblock \href {https://doi.org/10.1109/TSE.2006.31}
  {\path{doi:10.1109/TSE.2006.31}}.
\newline\urlprefix\url{http://ieeexplore.ieee.org/document/1610610/}

\bibitem{Mkaouer2016}
M.~W. Mkaouer, M.~Kessentini, S.~Bechikh, M.~{O Cinnéide}, K.~Deb,
  \href{http://link.springer.com/10.1007/s10664-015-9414-4}{{On the use of many
  quality attributes for software refactoring: a many-objective search-based
  software engineering approach}}, Empirical Software Engineering 21~(6) (2016)
  2503--2545.
\newblock \href {https://doi.org/10.1007/s10664-015-9414-4}
  {\path{doi:10.1007/s10664-015-9414-4}}.
\newline\urlprefix\url{http://link.springer.com/10.1007/s10664-015-9414-4}

\bibitem{Mohan2019}
M.~Mohan, D.~Greer,
  \href{https://linkinghub.elsevier.com/retrieve/pii/S0950584919300916}{{Using
  a many-objective approach to investigate automated refactoring}}, Information
  and Software Technology 112 (2019) 83--101.
\newblock \href {https://doi.org/10.1016/j.infsof.2019.04.009}
  {\path{doi:10.1016/j.infsof.2019.04.009}}.
\newline\urlprefix\url{https://linkinghub.elsevier.com/retrieve/pii/S0950584919300916}

\end{thebibliography}

\end{document}